\documentclass{article}
\usepackage[utf8]{inputenc}

\PassOptionsToPackage{numbers, compress}{natbib}
\usepackage[final]{neurips_2022}

\usepackage{amssymb}
\usepackage{amsmath}
\usepackage{mathtools}
\usepackage{physics}
\usepackage{xspace}
\usepackage{bm}
\usepackage{booktabs}
\usepackage{makecell}
\usepackage{relsize}
\usepackage[shortlabels,inline]{enumitem}
\usepackage{colortbl}
\usepackage{multirow}
\usepackage{xcolor}
\usepackage{flafter}
\usepackage{float}

\usepackage{tikz}
\usepackage{forest}
\usepackage{tikz-qtree}

\usepackage{linegoal}
\usepackage{amsthm}
\usepackage{thm-restate}
\usepackage[colorlinks]{hyperref}

\usepackage{algorithm}
\usepackage[noend]{algpseudocode}

\usepackage{cleveref}
\usepackage{autonum}

\makeatletter
\autonum@generatePatchedReferenceCSL{Cref}
\autonum@generatePatchedReferenceCSL{cref}
\makeatother

\makeatletter
\def\NAT@spacechar{~}%
\makeatother

\newcommand{\delimit}[3]{\newcommand{#1}[1]{\left#2##1\right#3}}
\delimit \ceil \lceil \rceil
\delimit \floor \lfloor \rfloor

\DeclareMathOperator*{\E}{\mathbb E}

\definecolor{darkgreen}{rgb}{0,0.667,0}
\hypersetup{citecolor=darkgreen}

\let\op\operatorname

\let\mc\mathcal

\newcommand{\R}{\ensuremath{\mathbb{R}}\xspace}
\renewcommand{\H}{\ensuremath{\mathcal{H}}\xspace}
\newcommand{\I}{\ensuremath{\mathcal{I}}\xspace}
\newcommand{\Z}{\ensuremath{\mathcal{Z}}\xspace}
\newcommand{\M}{\ensuremath{\mathcal{M}}\xspace}

\renewcommand{\S}{\ensuremath{\mathcal{S}}\xspace}

\newcommand{\X}{\ensuremath{\mathcal{X}}\xspace}

\newcommand{\mediator}{\textsf{\smaller M}}

\newcommand{\ra}{\textsc{rev}}
\newcommand{\rb}{\textsc{rec}}
\newcommand{\rc}{\textsc{act}}

\renewcommand{\vec}{\bm}

\newcommand{\Null}{\ensuremath{\varnothing}\xspace}
\newcommand{\chance}{{\sf C}}
\newcommand{\poly}{\op{poly}}
\newcommand{\ie}{{\em i.e.}\xspace}
\newcommand{\eg}{{\em e.g.}\xspace}

\theoremstyle{plain}
\newtheorem{theorem}{Theorem}[section]
\newtheorem{proposition}[theorem]{Proposition}

\theoremstyle{definition}
\newtheorem{definition}[theorem]{Definition}

\theoremstyle{remark}

\begingroup
    \makeatletter
    \@for\theoremstyle:=definition,remark,plain\do{%
        \expandafter\g@addto@macro\csname th@\theoremstyle\endcsname{%
            \setlength\thm@preskip\parskip
            \setlength\thm@postskip{0pt}
            }%
        }
\endgroup

\makeatletter
\renewenvironment{proof}[1][\proofname]{\par
  \pushQED{\qed}%
  \normalfont \topsep\z@skip %
  \trivlist
  \item[\hskip\labelsep
        \itshape
    #1\@addpunct{.}]\ignorespaces
}{%
  \popQED\endtrivlist\@endpefalse
}
\makeatother

\setlist{topsep=0pt}

\newcommand{\commentsymbol}{\it\color{gray}$\triangleright$~}
\algrenewcommand\algorithmiccomment[1]{\hfill{\commentsymbol#1}}

\newenvironment{ienumerate}{\begin{enumerate*}[{\bf 1)}]}{\end{enumerate*}}

\newcommand{\pone}{{\ensuremath{\color{p1color}\blacktriangle}}\xspace}
\newcommand{\ptwo}{{\ensuremath{\color{p2color}\blacktriangledown}}\xspace}
\definecolor{p1color}{RGB}{31,119,180}
\definecolor{p2color}{RGB}{255,127,14}
\definecolor{p3color}{RGB}{44,160,44}
\definecolor{p4color}{RGB}{214,39,40}

\usetikzlibrary{positioning}

\author{%
  Brian Hu Zhang \\
  Computer Science Department \\
  Carnegie Mellon University\\
  \texttt{bhzhang@cs.cmu.edu} \\
  \And
  Tuomas Sandholm \\
  Computer Science Department, CMU \\
  Strategic Machine, Inc.\\
  Strategy Robot, Inc.\\
  Optimized Markets, Inc.\\
  \texttt{sandholm@cs.cmu.edu} \\
}

\title{Polynomial-Time Optimal Equilibria with a Mediator in Extensive-Form Games}

\usepackage{caption}
\begin{document}

\maketitle

\begin{abstract}
    For common notions of correlated equilibrium in extensive-form games, computing an optimal (\eg, welfare-maximizing) equilibrium is NP-hard. Other equilibrium notions---{\em communication}~\cite{Forges86:Approach} and {\em certification}~\cite{Forges05:Communication} {\em equilibria}---augment the game with a mediator that has the power to both send and receive messages to and from the players---and, in particular, to remember the messages. In this paper, we investigate both notions in extensive-form games from a computational lens. We show that optimal equilibria in both notions can be computed in polynomial time, the latter under a natural additional assumption known in the literature. Our proof works by constructing a {\em mediator-augmented game} of polynomial size that explicitly represents the mediator's decisions and actions. Our framework allows us to define an entire family of equilibria by varying the mediator's information partition, the players' ability to lie, and the players' ability to deviate. From this perspective, we show that other notions of equilibrium, such as extensive-form correlated equilibrium, correspond to the mediator having {\em imperfect recall}. This shows that, at least among all these equilibrium notions, the hardness of computation is driven by the mediator's imperfect recall. 
    As special cases of our general construction, we recover 1) the polynomial-time algorithm of \citet{Conitzer04:Self} for {\em automated mechanism design} in Bayes-Nash equilibria and 2) the {\em correlation DAG} algorithm of \citet{Zhang22:Optimal} for optimal correlation.
    Our algorithm is especially scalable when the equilibrium notion is what we define as the {\em full-certification} equilibrium, where players cannot lie about their information but they can be silent. We back up our theoretical claims with experiments on a suite of standard benchmark games. 
\end{abstract}

\section{Introduction}

Various equilibrium notions in general-sum extensive-form games are used to describe situations where the players have access to a trusted third-party {\em mediator}, who can communicate with the players. Depending on the power of the mediator and the form of communication, these notions include the {\em normal-form}~\cite{Aumann74:Subjectivity} and {\em extensive-form} {\em correlated equilibrium} (NFCE and EFCE)~\cite{Stengel08:Extensive}, the {\em normal-form}~\cite{Moulin78:Strategically} and {\em extensive-form}~\cite{Farina20:Coarse} {\em coarse-correlated equilibrium} (NFCCE and EFCCE), the {\em communication equilibrium}~\cite{Forges86:Approach}, and the {\em certification equilibrium}~\cite{Forges05:Communication}. 

Several of these notions, in particular the EFCE and EFCCE, were defined for mainly {\em computational} reasons: the EFCE as a computationally-reasonable relaxation to NFCE, and the EFCCE as a computationally-reasonable relaxation of EFCE. When the goal is to compute a {\em single} correlated equilibrium, these relaxations are helpful: there are polynomial-time algorithms for computing an EFCE~\cite{Huang08:Computing}. However, from the perspective of computing {\em optimal} equilibria---that is, equilibria that maximize the expected value of a given function, such as the social welfare---even these relaxations fall short: for all of the {\em correlation} notions above, computing an optimal equilibrium of an extensive-form game is NP-hard~\cite{Stengel08:Extensive,Farina20:Coarse}.  

On the other hand, notions of equilibrium involving {\em communication} in games have arisen. These differ from the notions of {\em correlation} in that the mediator can receive and remember information from the players, and therefore pass information {\em between} players as necessary to back up their suggestions. {\em Certification equilibria}~\cite{Forges05:Communication} further strengthen communication equilibria by allowing players to {\em prove} certain information to the mediator.
To our knowledge, the computational complexity of optimal communication or certification equilibria has never been studied. We do so in this paper. The main technical result of our paper is a {\em polynomial-time algorithm} for computing optimal communication and certification equilibria (the latter under a certain natural condition about what messages the players can send). This stands in stark contrast to the notions of correlation discussed above.

To prove our main result, we define a general class of {\em mediator-augmented games}, each having polynomial size, that is sufficient to describe all of the above notions of equilibrium except the NFCE\footnote{We do not consider the NFCE, because it breaks our paradigm, which enforces that the mediator's recommendation be a single action. In NFCE, the whole strategy needs to be revealed upfront. It is an open question whether it is possible to even find {\em one} NFCE in polynomial time, not to mention an optimal one.}. We also build on this main result in several ways.
\begin{enumerate}
    \item We define the {\em full-certification} equilibrium, which is the special case in which players cannot lie to the mediator (but can opt out of revealing their information). In this case, the algorithm is a linear program whose size is {\em almost linear} in the size of the original game. As such, this special case scales extremely well compared to all of the other notions.
    \item We formalize notions for incorporating {\em payments} in the language of our augmented game. By using payments, mediators can incentivize players to play differently than they otherwise would, possibly to the benefit of the mediator's utility function.
    \item We define an entire family of equilibria using our augmented game, that includes as special cases the communication equilibrium, certification equilibrium, NFCCE, EFCCE, and EFCE. From this perspective, we show that other notions of equilibrium, such as extensive-form correlated equilibrium, correspond to the mediator having {\em imperfect recall}. This shows that, at least among all these equilibrium notions, the hardness of computation is driven by the mediator's imperfect recall. 
    We argue that, for this reason, many stated practical applications of correlated equilibria should actually be using communication or certification equilibria instead, which are both easier to compute (in theory, at least) and better at modelling the decision-making process of a rational mediator.
    \item We empirically verify the above claims via experiments on a standard set of game instances.
\end{enumerate}

{\bf Applications and related work.}
Correlated and communication equilibria have various applications that have been well-documented. Here, we discuss just a few of them, as motivation for our paper. For further discussion of related work, especially relating to automated dynamic mechanism design and persuasion, see \Cref{se:related}.

{\it Bargaining, negotiation, and conflict resolution}~\cite{Chalamish12:Automed,Farina19:Correlation}. Two parties with asymmetric information wish to arrive at an agreement, say, the price of an item. A mediator, such as a central third-party marketplace, does not know the players' information but can communicate with the players.

{\it Crowdsourcing and ridesharing}~\cite{Furuhata13:Ridesharing,Ma21:Spatio,Zhang22:Optimal}. A group of players each has individual goals (\eg, to make money by serving customers at specific locations). The players are coordinated by a central party (\eg, a ridesharing company) that has more information than any one of the players, but the players are free to ignore recommendations if they so choose. 

{\it Persuasion in games}~\cite{Kamenica11:Bayesian,Celli20:Private,Mansour22:Bayesian,Gan22:Bayesian,Wu22:Sequential}. The mediator (in that literature, usually ``sender'') has more information than the players (``receivers''), and wishes to tell information to the receivers so as to persuade them to act in a certain way.

{\it Automated mechanism design}~\cite{Conitzer02:Complexity,Conitzer04:Self,Zhang21:Automated,Zhang22:Planning,Papadimitriou22:Complexity,Zhang21:Automatedb,Kephart15:Complexity,Kephart21:Revelation}.  Players have private information unknown to the mediator. The mediator wishes to commit to a strategy---that is, set a mechanism---such that players are incentivized to honestly reveal their information. In fact, in \Cref{se:payments} we will see that we recover the polynomial-time Bayes-Nash randomized mechanism design algorithm of \citep{Conitzer02:Complexity,Conitzer04:Self} as a special case of our main result.

Some of the above examples are often used to motivate correlated equilibria. However, when the mediator is a rational agent with the ability to remember information that it is told and pass the information between players as necessary, we will argue that communication or certification equilibrium should be the notion of choice, for both conceptual and computational reasons.

\section{Preliminaries}
In this section, we discuss background on correlation in extensive-form games. 

{\bf Extensive-form games.} An {\em extensive-form game} $\Gamma$ with $n$ players consists of the following.
\begin{enumerate}
    \item A directed tree of {\em nodes} or {\em histories} \H, whose root is denoted \Null. The depth of the tree will be denoted $T$. The edges out of nodes are labeled with {\em actions}, and the set of such actions will be denoted $A_h$. Given a node $h \in H$ and action $a$ at $h$, the child reached by following action $a$ at node $h$ is denoted $ha$. The set of terminal (leaf) nodes in \H is denoted \Z. Terminal nodes will always be denoted $z$ throughout the paper.
    \item A partition $\H \setminus \Z = \H_\chance \sqcup \H_1 \sqcup \dots \sqcup \dots \H_n$ of nodes, where $\H_i$ is the set of all nodes at which player $i$ plays and player $\H_\chance$ is the set of chance nodes.
    \item For each player $i$, a partition $\I_i$ of player $i$'s decision nodes, $\H_i$, into {\em information sets} or {\em infosets}. Every node in a given information set $I$ must have the same set of actions, denoted $A_I$. We will call the partition $\I = \I_1 \sqcup \dots \sqcup \I_n$ the {\em players' information partition}.
    \item For each player $i$, a {\em utility vector} $\vec u_i \in [0,1]^\Z$, where $u_i[z]$ denotes the utility achieved by player $i$ at terminal node $z$.
    \item For each chance node $h \in \H_\chance$, a probability distribution $p(\cdot|h)$ over the children of $h$.
\end{enumerate}

The {\em sequence} $\sigma_i(h)$ is the list of infosets reached by player $i$, and actions taken by the player $i$ at those infosets, on the $\Null \to h$ path, {\em not} including the infoset at $h$ itself (if any). 
We will assume that each player has {\em perfect recall}---that is, for each infoset $I$, the sequence of the player acting at $I$ should be the same for each node in $I$. We will denote this sequence $\sigma(I)$. In perfect-recall games, nonempty sequences will be identified by the last infoset-action pair $Ia$ in them.

We also will assume that games are {\em timeable} and {\em fixed-turn-order}, that is, information sets do not span multiple levels of the tree, and all nodes in the same layer of the tree belong to the same player\footnote{Timeability is not without loss of generality, but any game for which the precedence order $\preceq$ is a partial order over infosets can be converted to a timeable game by adding dummy nodes. Given timeability, fixed-turn-order is without loss of generality, also by adding dummy nodes}.

We will use the following notation. The relation $\preceq$ denotes the natural precedence order induced by the tree \H: we write $h \preceq h'$ means that $h$ is an ancestor of $h'$ (or $h = h'$), and for sets $S, S'$, we say $S \preceq S'$ if there are some $h \in S, h' \in S'$ such that $h \preceq h'$.  The binary operation $\land$ denotes the lowest common ancestor: $h \land h'$ is the lowest node $u$ such that $u \preceq h, h'$.  %

For sequences, $\vec\sigma(h) = (\sigma_1(h), \dots, \sigma_n(h))$ denotes the {\em joint sequence} of all players at node $h$. $N(\sigma)$ denotes the set of possible {\em next} infosets following sequence $\sigma$, that is, $N(\sigma) = \{ I : \sigma(I) = \sigma\}$. The set $\Sigma_i$ denotes the set of sequences of player $i$, and $\Sigma$ denotes the set of all sequences across all players (i.e., $\Sigma = \sqcup_i \Sigma_i$). 

A {\em pure strategy} for a player $i$ is a selection of one action for each information set $I \in \I_i$. A {\em pure profile} is a tuple of pure strategies. A {\em correlated profile} is a distribution over pure profiles. 

We will generally work with strategies in {\em realization form} (see e.g., \citet{Koller94:Fast}). Given a pure strategy $\vec x$, we say that $\vec x$ {\em plays to} $z \in \mc Z$ if $\vec x$ plays every action on the $\Null \to z$ path. We will call the vector $\vec x \in \{0, 1\}^{\mc Z}$ the {\em realization form} of $\vec x$. The realization form of a mixed strategy is the appropriate convex combination. The set of mixed strategies forms a convex subset of $\R^{\mc Z}$ that, so long as the player has perfect recall, can be expressed using linearly many constraints and variables.

We will occasionally need to discuss changing information partitions of $\Gamma$. If $\mc J = \mc J_1 \sqcup \dots \sqcup \mc J_n$ is another valid information partition, we will use $\Gamma^{\mc J}$ to denote the game $\Gamma$ with its information partition replaced by $\mc J$. We will also occasionally need to talk about multiple games simultaneously; where this is the case, we will mark attributes of the game the same as the game itself. For example, $\hat\H$ is the node set of game $\hat\Gamma$. 

{\bf Communication and certification equilibria.}
Here, we review definitions related to {\em communication equilibria}, following \citet{Forges86:Approach} and later related papers.

\begin{definition}
Let $S$ be a space of possible {\em messages}. A {\em pure mediator strategy}
is a map $d : S^{\le T} \to S$, where $S^{\le T}$ denotes the set of sequences in $S$ of length at most $T$. A {\em randomized mediator strategy} (hereafter simply {\em  mediator strategy}) is a distribution over pure mediator strategies. 
\end{definition}
We will assume that the space of possible messages is large, but not exponentially so. In particular, we will assume that $\{\bot\} \cup \mc I \cup \bigcup_h A_h \subseteq S$ (\ie, messages can at least be nothing, information, or actions)\footnote{{\em A priori}, although the messages are given these names, they carry no semantic meaning. The revelation principle  is used to assign natural meaning to the messages.}  and that $\abs{S} \le \poly(\abs{\H})$. The latter assumption is mostly for cleanliness in stating results: we will give algorithms that need $S$ as an input that we wish to run in time $\poly(\abs{\H})$.

A mediator strategy augments a game as follows. If the strategy is randomized, it first samples a pure strategy $d$, which is hidden from the players. At each timestep $t$, a player reaches a history $h$ at which she must act, and observes the infoset $I \ni h$. She sends a message $s_t \in S$ to the mediator. The mediator then sends a response $d(s_1, \dots, s_t)$, which depends on the message $s_t$ as well as the messages sent by all other players prior to timestep $t$. Then, the player chooses her action $a \in A_h$. We will call the sequence of messages sent and received between the mediator and player $i$, the {\em transcript with player $i$}.
A {\em communication equilibrium}\footnote{Previous models of communication in games~\cite{Forges86:Approach,Forges05:Communication} usually worked with a model in which players send messages, receive messages, and play moves {\em simultaneously}, rather than in sequence as in the extensive-game model that we use. The simultaneous-move model is easy to recreate in extensive form: by adding further ``dummy nodes'' at which players learn information but only have one legal action, we can effectively re-order when players ought to communicate their information to the mediator.} is a Nash equilibrium of the game $\Gamma$ augmented with a mediator strategy.
The mediator is allowed to perform arbitrary communication with the players. In particular, the mediator is allowed to {\em pass information from one player to another}. Further, the players are free to send whatever messages they wish to the mediator, including false or empty messages. These two factors distinguish communication equilibria from notions of {\em correlated equilibria}. In \Cref{se:family} we will discuss this comparison in greater detail. 

A useful property in the literature on communication equilibria is the {\em revelation principle} (\eg, \cite{Forges86:Approach}). Informally, the revelation principle states that any outcome achievable by an {\em arbitrary} strategy profile can also be achieved by a {\em direct} strategy profile, in which the players tell the mediator all their information and are subsequently directly told by the mediator which action to play.
In order to be a communication equilibrium, the players still must not have any incentive to deviate from the protocol. That is, the equilibrium must be {\em robust} to all messages that a player may attempt to send to the mediator, even if {\em in equilibrium} the player always sends the honest message.

\citet{Forges05:Communication} further introduced a form of equilibrium for Bayesian games which they called {\em certification equilibria}. In certification equilibria, the messages that a player may legally send are dependent on their information; as such, some messages that a player can send are {\em verifiable}. 
At each information set $I \in \I$, let $S_I \subseteq S$ denote the set of messages that the player at infoset $I$ may send to the mediator. We will always assume that $I \in S_I$ and $\bot \in S_I$ for all $I$. That is, all players always have the options of revealing their true information or revealing nothing.

\section{Extensive-form \S-certification equilibria}\label{se:proof}

The central notion of interest in this paper is a generalization of the notion of certification equilibria \cite{Forges05:Communication} to extensive-form games. 

\begin{definition}
Given an extensive-form game $\Gamma$ and a family of valid message sets $\S = \{ S_I : I \in \I\}$, an {\em \S-certification equilibrium} is a Nash equilibrium of the game augmented by a randomized mediator, in which each player at each information set $I$ is restricted to sending a message $s \in S_I$. 
\end{definition}

The existence of \S-certification equilibria follows from the existence of {\em Nash} equilibria, which are the special case where the mediator does nothing.

We will need one extra condition on the message sets, which is known as the {\em nested range condition} (NRC)~\cite{Green77:Characterization}: if $I \in S_{I'}$, then $S_{I} \subseteq S_{I'}$. That is, if a player with information $I'$ can lie by pretending to have information $I$, then that player can also emulate any other message she would have been able to send at $I$. Equivalently, the honest message $I$ should be the {\em most certifiable} message that a player can send at infoset $I$.
Our main result is the following.

\begin{theorem}\label{th:main}
Let $\vec u_\mediator \in \R^\Z$ be an arbitrary utility vector for the mediator. Then there is a polynomial-time algorithm that, given a game $\Gamma$ and a message set family \S satisfying the nested range condition, computes an optimal \S-certification equilibrium, that is, one that maximizes $\E_z u_\mediator[z]$ where the expectation is over playouts of the game under equilibrium.
\end{theorem}

In particular, by setting $S_I = S$ for all $I$, \Cref{th:main} implies that optimal communication equilibria can be computed in polynomial time.

The rest of the paper is organized as follows. First, we will prove our main theorem. Along the way, we will demonstrate a form of revelation principle for \S-certification equilibria. We will then discuss comparisons to other known forms of equilibrium, including the extensive-form correlated equilibrium~\cite{Stengel08:Extensive}, and several other natural extensions of our model. Finally, we will show experimental results that compare the computational efficiency and social welfare of various notions of equilibrium on some experimental game instances.

\subsection{Proof of \Cref{th:main}: The single-deviator mediator-augmented game}\label{se:proof-main}
In this section, we construct a game $\hat\Gamma$, with $n+1$ players, that describes the game $\Gamma$ where the mediator has been added as an explicit player. This game has similar structure to the one used by \citet[Corollary 2]{Forges86:Approach}, but, critically, has size polynomial in $\abs{\H}$. This is due to two critical differences. First, the players are assumed to either send $\bot$, or send messages that mediator cannot immediately prove to be off-equilibrium. In particular, if the player's last message was $I$ and the mediator recommended action $a$ at $I$, the player must send a message $I'$ with $\sigma(I') = Ia$. If this is impossible, the player must send $\bot$. Therefore, in particular, we will assume that $S_I$ consists of only $\bot$ and information sets $I'$ at the same level as $I$. Second, only one player is allowed to deviate. Therefore, the strategy of the mediator is not defined in cases where two or more players deviate.

We now formalize $\hat\Gamma$. Nodes in $\hat\Gamma$ will be identified by tuples $(h, \vec\tau, r)$ where $h \in \H$ is a history in $\Gamma$, $\vec\tau = (\tau_1, \dots, \tau_n)$ is the collection of transcripts with all players, and $r \in \{\ra, \rb, \rc\}$ is a {\em stage marker} that denotes whether the current state is one in which a player should be {\em revealing information} (\ra), the mediator should be {\em recommending a move} (\rb), or the player should be {\em selecting an action} (\rc).
The progression of $\hat\Gamma$ is then defined as follows. We will use the notation $\vec\tau[i{\cdot}s]$ to denote appending message $s$ to $\tau_i$. 

\begin{itemize}
    \item The root node of $\hat\Gamma$ is $(\Null, (\Null, \dots, \Null), \ra)$. 
    \item Nodes $(z, \vec\tau, \ra)$ for $z \in \Z$ are also terminal in $\Gamma$. The mediator gets utility $u_\mediator[z]$, where $u$ is the mediator's utility function as in \Cref{th:main}. All other players $i$ get utility $u_i[z]$.  
    \item Nodes $(h, \vec\tau, \ra)$ for non-terminal $h$ are decision nodes for the player $i$ who acts at $h$.
    \begin{enumerate}
        \item If $i$ is chance, there is one valid transition, to $(h, \vec\tau, \rc)$.
        \item  If some other player $j \ne i$ has already deviated (i.e., $\sigma_j(h) \ne \tau_j)$), there is one valid transition, to $(h, \vec\tau[i{\cdot}I], \rb)$ where $I \ni h$.
        \item If player $i$ has deviated or no one has deviated, then player $i$ observes the infoset $I \ni h$, and selects a legal message $I' \in S_I \cap (\{ \bot\} \cup N(\tau_i))$ to send to the mediator\footnote{If $\tau_i$ contains any $\bot$ messages, then we take $N(\tau_i) = \varnothing$}. Transition to $(h, \vec\tau[i{\cdot}I'], \ra)$.
    \end{enumerate}
    \item At $(h, \vec\tau, \rb)$ where $h \in \H_i$, the mediator observes the transcript $\tau_i$ and makes a {\em recommendation} $a$. If $\tau_i$ contains any $\bot$ messages, then $a = \bot$. Otherwise, $a$ is a legal action $a \in A_I$, where $I$ is the most recent message in $\tau_i$. Transition to $(h, \vec\tau[i{\cdot}a], \rc)$. 
    \item Nodes $(h, \vec\tau, \rc)$ for non-terminal $h$ are decision nodes for the player $i$ who acts at $h$.
    \begin{enumerate}
        \item If $i$ is chance, then chance samples a random action $a \sim p(\cdot|h)$. Transition to $(ha, \vec\tau, \ra)$.
        \item  If some other player $j \ne i$ has already deviated, there is one valid transition, to $(ha, \vec\tau, \rb)$, where $a$ is the action sent by the mediator.
        \item If player $i$ has deviated or no one has deviated, then player $i$ observes the transcript $\tau_i$, and selects an action $a' \in A_h$. Transition to $(ha', \vec\tau, \ra)$. The action $a'$ need not be the recommended action.
    \end{enumerate}
\end{itemize}

Since at most one player can ever deviate by construction, and the length of the transcripts are fixed because turn order is common knowledge, the transcripts $\vec\tau$ can be identified with {\em sequences} ${\sigma_i}$ of the deviated player, if any. We will make this identification: we will use the shorthand $h^{\sigma_i}$ to denote the history $(h, (\sigma_{-i}(h), \sigma_i), \ra)$, and $h^\bot$ for $(h, \vec\sigma(h), \ra)$ (i.e., no one has deviated yet). Therefore, in particular, this game has at most $O(\abs{\H} \abs{\Sigma})$ histories. 

For each non-mediator player, there is a well-defined {\em direct strategy} $\vec{\hat x}^*_i$ for that player: always report her true information $I \ni h$, and always play the action recommended by the mediator. The goal of the mediator is to {\em find a strategy $\vec{\hat x}_\mediator$ for itself that maximizes its expected utility, subject to the constraint that each player's direct strategy is a best response}---that is, find $\vec{\hat x}_\mediator$ such that $(\vec{\hat x}_\mediator, \vec{\hat x}^*_1, \dots, \vec{\hat x}^*_n)$ is a (strong) Stackelberg equilibrium of $\hat\Gamma$.

We claim that finding a mediator strategy $\vec{\hat x}_\mediator$ that is a strong Stackelberg equilibrium in $\hat\Gamma$ is equivalent to finding an optimal \S-certification equilibrium in $\Gamma$. We prove this in two parts. First, we prove a version of the revelation principle for \S-certification equilibria.
\begin{definition}
An \S-certification equilibrium is {\em direct} if it satisfies the following two properties.
\begin{enumerate}
    \item ({\em Mediator directness}) If the transcript $\tau_i$ of a player $i$ is exactly some sequence of player $i$, and player $i$ sends an infoset $I$ with $\sigma(I) = \tau_i$, then the mediator replies with an action $a \in A_I$. Otherwise\footnote{This condition is necessary because, if the mediator does not know what infoset the player is in, the mediator may not be {\em able} to send the player a valid action, because action sets may differ by infoset. }, the mediator replies $\bot$.
    \item ({\em Player directness}) In equilibrium, players always send their true information $I$, and, upon receiving an action $a \in A_I$, always play that action.
\end{enumerate}
\end{definition}

\begin{proposition}[Revelation principle for \S-certification equilibria under NRC]\label{pr:rev}
Assume that \S satisfies the nested range condition. For any \S-certification equilibrium, there is a realization-equivalent direct equilibrium.
\end{proposition}

Omitted proofs can be found in the appendix.
Since direct mediator strategies are exactly the mediator strategies in $\hat\Gamma$, and the player strategies are only limited versions of what they are allowed to do in \S-certification equilibrium, this implies that, for any \S-certification equilibrium, there is a mediator strategy $\vec{\hat x}_\mediator$ in $\hat\Gamma$ such that $(\vec{\hat x}_\mediator, \vec{\hat x}^*_1, \dots, \vec{\hat x}^*_n)$ is a Stackelberg equilibrium. We will also need the converse of this statement.

\begin{proposition}\label{pr:rev2}
Let $\vec{\hat x}_\mediator$ be a strategy for the mediator in $\hat\Gamma$ such that, in the strategy profile $(\vec{\hat x}_\mediator, \vec{\hat x}^*_1, \dots, \vec{\hat x}^*_n)$, every $\vec{\hat x}^*_i$ for $i \ne \mediator$ is a best response. Then there is an direct \S-certification equilibrium that is realization-equivalent to $(\vec{\hat x}_\mediator, \vec{\hat x}^*_1, \dots, \vec{\hat x}^*_n)$.
\end{proposition}

Therefore, we have shown that the mediator strategies $\vec{\hat x}_\mediator$ in $\hat\Gamma$ for which $(\vec{\hat x}_\mediator, \vec{\hat x}^*_1, \dots, \vec{\hat x}^*_n)$ is a Stackelberg equilibrium in $\hat\Gamma$ correspond exactly to optimal \S-certification equilibria of $\Gamma$. Such a Stackelberg equilibrium can be found by solving the following program:
\begin{align}
\begin{aligned}
    \max_{\vec{\hat x}_\mediator \in \hat\X_\mediator} \quad& \sum_{\hat z \in \hat\Z} \hat x_\mediator[\hat z] \hat u_\mediator[\hat z] \hat p(\hat z) \prod_{i \in [n]} \hat x^*_i[\hat z]\\
    \text{s.t.} \quad& \max_{\vec{\hat x}'_j \in \hat{\mc X}_j} \sum_{\hat z \in \hat\Z} \hat x_\mediator[\hat z] \hat u_i[\hat z] \hat p(\hat z) \qty(\hat x_{j}'[\hat z] - \hat x_j^*[\hat z]) \prod_{i \ne j} \hat x_i^*[\hat z] \le 0\quad \forall{j \in [n]}
\end{aligned} \label{eq:program}
\end{align}
where $\hat{\mc X}_i$ is the sequence-form strategy space~\cite{Koller94:Fast} of player $i$ in $\hat\Gamma$.

The only variables in the program are $\vec{\hat x}_i$ for each player $i$ and the mediator. In particular, the direct strategies $\hat{\vec x}_i^*$ are constants. Therefore, the objective is a linear function, and the inner maximization constraints are bilinear in $\hat{\vec x}_\mediator$ and $\hat{\vec x}_j$. Therefore, this program can be converted to a linear program by dualizing the inner optimizations. For more details on this conversion, see \Cref{se:lp-full}. The result is a linear program of size $O(n\abs*{\hat\H}) = O(n\abs{\H} \abs{\Sigma})$. We have thus proved \Cref{th:main}.

\subsection{Extensions and special cases}\label{se:extensions}
In this section, we describe several extensions and interesting special cases of our main result.

{\bf Full-certification equilibria.}
One particular special case of \S-certification equilibria which is particularly useful. We define a {\em full-certification equilibrium} as an \S-certification equilibrium where $S_I = \{ \bot, I \}$. Intuitively, this means that players cannot {\em lie} to the mediator, but they may {\em withhold} information. We will call such an equilibrium {\em full-certification}.
Removing valid messages from the players only reduces their ability to deviate and thus increases the space of possible equilibrium strategies. As such, the full-certification equilibria are the largest class of \S-certification equilibria. 

For full-certification equilibria, the size of game $\hat\Gamma$ reduces dramatically. Indeed, in all histories $h^{Ia}$ of $\hat\Gamma$, we must have $I \preceq h$. Therefore, we have $|\hat\H| \le \abs{\H} BD$ where $B$ is the maximum branching factor and $D$ is the depth of the game tree, \ie, the size of $\hat\Gamma$ goes from essentially quadratic to essentially quasilinear in $\abs{\H}$. The mediator's decision points in $\hat\Gamma$ for a full-certification equilibrium are the {\em trigger histories} used by \citet{Zhang22:Optimal} in their analysis of various notions of correlated equilibria. Later, we will draw further connections between full certification and correlation.

{\bf Changing the mediator's information.}
In certain cases, the mediator, in addition to messages that it is sent by the players, also has its own observations about the world. These are trivial to incorporate into our model: simply change the information partition of the mediator in $\hat\Gamma$ as needed. Alternatively, one can imagine adding a ``player'', with no rewards (hence no incentive to deviate), whose sole purpose is to observe information and pass it to the mediator. For purposes of keeping the game small, it is easier to adopt the former method. To this end, consider any refinement partition $\M$ of the mediator infosets in $\hat\Gamma$, and consider the game $\hat\Gamma^\M$ created by replacing the mediator's information partition in $\hat\Gamma$ with $\M$. Then we make the following definition.
\begin{definition}
An {\em $(\S, \M)$-certification equilibrium} of $\Gamma$ is a mediator strategy $\hat x_\mediator$ in $\hat\Gamma^\M$ such that, in the strategy profile $(\hat x_\mediator, \hat x_1^*, \dots, \hat x_n^*)$, every $x^*_i$ for $i \ne \mediator$ is a best response.
\end{definition}

$(\S, \M)$-certification equilibria may not exist: indeed, if $\M$ is coarser than the mediator's original information partition in $\hat\Gamma$, then the mediator may not have enough information to provide good recommendations under the restrictions of $\hat\Gamma$. This can be remedied by allowing payments (see \Cref{se:payments}), or by making the assumption that the mediator {\em at least} knows the transcript of the player to whom she is making any nontrivial recommendation: 

\begin{definition}
A mediator partition \M is {\em direct} if, at every mediator decision point $(h, \vec\tau, \rb)$, so long as $\abs{A_h} > 1$, the mediator knows the transcript of the player acting at $h$. $\mc M$ is {\em strongly direct} if the mediator also observes the transcript when $\abs{A_h} = 1$. 
\end{definition}
The condition $\abs{A_h} > 1$ in the definition allows the mediator to possibly {\em not} observe the full information of a player if she does not need to make a nontrivial recommendation to that player. In particular, this allows players to sometimes have information that they only partially reveal to the mediator, so long as the player does not immediately need to act on such information. 

{\bf Coarseness.} In literature on correlation, {\em coarseness} refers to the restriction that a player must obey any recommendation that she receives (but may choose to deviate by not requesting a recommendation and instead playing any other action). {\em Normal-form coarseness} further adds the restriction that players can only choose to deviate at the start of the game---the mediator essentially takes over and plays the game on behalf of non-deviating players. These notions can easily be expressed in terms of our augmented game, therefore also allowing us to express coarse versions of our equilibrium notions as augmented games. %

\subsection{The gap between polynomial and not polynomial}

If players cannot send messages to the mediator at all, and the mediator has no other way of gaining any information, we recover the notion of {\em autonomous correlated equilibrium (ACE)}. It is NP-hard to compute optimal ACE, even in Bayesian games (see \eg, \citet{Stengel08:Extensive}).

When $\mc M$ is direct and perfect recall, computing an optimal {\em direct} $(\mc S, \mc M)$-certification equilibrium can be done in polynomial time using our framework. When $\mc S$ obeys NRC and $\mc M$ satisfies a stronger condition\footnote{Roughly speaking, this condition is that players should not be able to cause the mediator to gain information apart from their own messages by sending messages. It holds for all notions we discuss in this paper. Formalizing the general case is beyond the scope of this paper. }, the proof of the revelation principle (\Cref{pr:rev,pr:rev2}) works, and the resulting equilibrium is guaranteed to be optimal over all possible equilibria including those that may not be direct.

If NRC does not hold, one can still solve the program \eqref{eq:program}, and the solution is still guaranteed to be an optimal {\em direct} equilibrium by \Cref{pr:rev2}. However, it is not guaranteed to be optimal over all possible communication structures. Indeed, \citet[Theorem 1]{Green77:Characterization} give an instance in which, without NRC, there can be an outcome distribution that is not implementable by a direct mediator. Our program cannot find such an outcome distribution. The counterexample does not preclude the possibility of efficient algorithms for finding optimal certification equilibria in more general cases, but does give intuition for why NRC is crucial to our construction.

We could also consider changing the mediator's information partition so that the mediator does not have perfect recall. This transformation allows us to recover notions of {\em correlation} in games. Indeed, if we start from the {\em full-certification} equilibrium and only allow the mediator to remember the transcript with the player she is currently talking to, we recover EFCE. Adding coarseness similarly recovers EFCCE and NFCCE. In this setting, the inability to represent the strategy space of an imperfect-recall player may result in the loss of efficient algorithms.

\subsection{A family of equilibria}\label{se:family}

By varying 
\begin{ienumerate}
    \item what the mediator observes,
    \item whether the mediator has perfect recall, 
    \item whether the players can lie or only withhold information, and
    \item when and how players can deviate from the mediator's recommended actions,
\end{ienumerate}
we can use our framework to define a family consisting of 16 conceptually different 
equilibrium notions. More can be generated by considering other variations in this design space, but we focus on the extreme cases in the table. Some of these were already defined in the literature; the remaining names are ours. The result is \Cref{ta:notions}. An inclusion diagram for these notions can be found in \Cref{se:inclusions}.

In the table, {\em ex ante} means that players have only a binary choice between deviating (in which case they can play whatever they want) and playing (in which case they must always be direct and obey recommendations). With \textit{ex ante} deviations, it does not matter whether lying is allowed because we can never get to that stage: either the player deviates immediately and never communicates with the mediator, or the player is direct. If the mediator only remembers the current active player's information, and players cannot lie, withholding and coarsely deviating are the same. 

{\em Mediator information advantage} means that the mediator always learns the infoset of the current active player, and therefore requires no messages from the players. This is equivalent to forcing players to truthfully report information. A mediator with information advantage may still not have perfect information---for example, it will not know whether a player (or nature) has played an action until some other player observes the action. In this setting, the mediator may also have extra private information (known to none of the players), leading to the setting of Bayesian persuasion~\cite{Kamenica11:Bayesian}. In extensive-form games, there are two different reasonable notions of persuasion: one that stems from extending {\em correlated} equilibria, and one that stems from extending {\em communication} equilibria. The distinction is that, in the former, the mediator has imperfect recall. For a more in-depth discussion of Bayesian persuasion, see \Cref{se:related}.

Our framework allows optimal equilibria for all notions in the table to be computed. For perfect-recall mediators, this is possible in polynomial time via the sequence form; for imperfect-recall mediators, the problem is NP-hard, in general, but the {\em team belief DAG} of \citet{Zhang22:Team_DAG} can be used to recover fixed-parameter algorithms. For the notions of correlated equilibrium, this method results in basically the same LP as the {\em correlation DAG} of \citet{Zhang22:Optimal}.

We do not claim that all of these notions are easy to motivate. For example, correlated equilibria are usually arrived at in the ``truth known, imperfect recall'' setting; the correlated equilibrium notions where lying is allowed are more difficult to motivate in this respect. Further, even the fixed-parameter algorithms of \citet{Zhang22:Optimal} would fail in this setting, because ``public states'' can no longer be treated as public due to the possibility of lying players. We leave to future research the problem of finding a motivation for the notions that we do not reference elsewhere in the paper.

\begin{table}[!tb]
\newcommand{\mymidrule}[1]{\arrayrulecolor{lightgray} \cmidrule{2-2} \cmidrule{#1-5} \arrayrulecolor{black}}
    \caption{A whole family of equilibria. See \Cref{se:family} for an explanation of the terms used in the table. {\em NF, EF}, and {\em IR} stand for {\em normal-form}, {\em extensive-form}, and {\em imperfect-recall} respectively.}
    \centering\tiny
    \setlength{\tabcolsep}{1mm}
    \begin{tabular}{cc|cccc}
    && \multicolumn{3}{c}{\bf when can players deviate?} \\
    &&\bf ex ante & \multicolumn{2}{c}{\bf ex interim} \\
    && & \bf coarse & \bf not coarse \\
    \midrule
    \multirow{3}{*}{\makecell{\bf mediator remembers only \\ \bf  current player's transcript}} & \bf lying possible & \multirow{2}{*}{\makecell{NFCCE~\cite{Moulin78:Strategically}}} & truthful EFCCE & truthful EFCE \\\mymidrule4
    &\bf  withholding only & & EFCCE~\cite{Farina20:Coarse} & EFCE~\cite{Stengel08:Extensive}
    \\\mymidrule3
& \bf mediator information advantage & NF coarse IR persuasion~\cite{Celli20:Private} & coarse IR persuasion & IR persuasion
\\\midrule
     &\bf  lying possible & \multirow{2}{*}{\makecell{NF coarse full-cert \\(``mediated''~\cite{Monderer09:Strong})}} & coarse comm & comm~\cite{Forges86:Approach}\\\mymidrule4
\bf \multirow{2}{*}{mediator perfect recall} &\bf withholding only & & coarse full-cert & full-cert~\cite{Forges05:Communication} \\\mymidrule3
& \bf mediator information advantage & NF coarse persuasion & coarse persuasion & persuasion
    \end{tabular}
    \label{ta:notions}
\end{table}

\section{Experiments}
\begin{table}[!tb]
\setlength{\fboxsep}{2pt}
\newcommand{\cbox}[2]{\fcolorbox{white}[rgb]{#1}{\phantom{56m 31s}\llap{#2}}}
\newcommand{\mymidrulegray}{\arrayrulecolor{lightgray}\mymidrule}
\newcommand{\mymidruledgray}{\arrayrulecolor{gray}\mymidrule}
\newcommand{\mymidrule}{
\midrule
\arrayrulecolor{black}}
\newcommand{\unk}{\textcolor{black!30}{---}}
\setlength{\tabcolsep}{.5mm}
\newcommand{\tworow}[1]{\multirow{2}{*}{#1}}
\centering
\caption{Table of experimental results. Values are the optimal social welfare given the type of equilibrium. Values and timings for optimal correlated equilibria were taken from \citet{Zhang22:Optimal} and are included here for purposes of comparison. When payments are used, the mediator is informed before making the payment of whether the player was honest, and the optimization objective is the social welfare of the original terminal state, minus any payments made. ``oom'' is out of memory.}
{\tiny
\begin{tabular}{lrc|ccc|cccccccc}
&&& \multicolumn{3}{c|}{[ZFCS'22]} &  \multicolumn{8}{c}{\it This paper} \\
\bf game & \bf $\abs{\Z}$ & & \bf NFCCE & \bf EFCCE & \bf EFCE & \multicolumn{2}{c}{\bf NF Coarse Cert} & \multicolumn{2}{c}{\bf Coarse Cert} & \multicolumn{2}{c}{\bf Cert} & \multicolumn{2}{c}{\bf Comm} \\
&&& &&& no pay & pay & no pay & pay & no pay & pay & no pay & pay
\\\midrule
\tworow{ B222 } & \tworow{ 1072 } & value & \cbox{0.7843137254901961,0.7843137254901961,1.0}{0.000} & \cbox{0.9649365628604383,0.8778162245290273,0.8778162245290273}{-0.525} & \cbox{0.9649365628604383,0.8778162245290273,0.8778162245290273}{-0.525} & \cbox{0.7843137254901961,0.7843137254901961,1.0}{0.000} & \cbox{0.7843137254901961,0.7843137254901961,1.0}{0.000} & \cbox{0.9649365628604383,0.8778162245290273,0.8778162245290273}{-0.525} & \cbox{0.9233371780084583,0.9233371780084583,0.9478662053056517}{-0.333} & \cbox{0.9649365628604383,0.8778162245290273,0.8778162245290273}{-0.525} & \cbox{0.9534025374855825,0.9085736255286428,0.9085736255286428}{-0.453} & \cbox{1.0,0.7843137254901961,0.7843137254901961}{-0.750} & \cbox{0.9640138408304498,0.8802768166089965,0.8802768166089965}{-0.520} \\
&& time &  \cbox{0.8310649750096116,0.8310649750096116,0.9824682814302191}{0.02s} & \cbox{0.8938100730488273,0.8938100730488273,0.9589388696655132}{0.05s} & \cbox{0.9547866205305652,0.904882737408689,0.904882737408689}{0.17s} & \cbox{0.7843137254901961,0.7843137254901961,1.0}{0.01s} & \cbox{0.7843137254901961,0.7843137254901961,1.0}{0.01s} & \cbox{0.8310649750096116,0.8310649750096116,0.9824682814302191}{0.02s} & \cbox{0.9171856978085352,0.9171856978085352,0.9501730103806229}{0.07s} & \cbox{0.9061130334486736,0.9061130334486736,0.9543252595155709}{0.06s} & \cbox{0.9547866205305652,0.904882737408689,0.904882737408689}{0.17s} & \cbox{1.0,0.7843137254901961,0.7843137254901961}{3.80s} & \cbox{1.0,0.7843137254901961,0.7843137254901961}{4.05s} \\\mymidrulegray
\tworow{ B322 } & \tworow{ 19116 } & value & \cbox{0.7843137254901961,0.7843137254901961,1.0}{0.000} & \cbox{1.0,0.7843137254901961,0.7843137254901961}{-0.317} & \cbox{1.0,0.7843137254901961,0.7843137254901961}{-0.317} & \cbox{0.7843137254901961,0.7843137254901961,1.0}{-0.000} & \cbox{0.7843137254901961,0.7843137254901961,1.0}{0.000} & \cbox{1.0,0.7843137254901961,0.7843137254901961}{-0.317} & \cbox{0.9566320645905421,0.8999615532487505,0.8999615532487505}{-0.200} & \cbox{1.0,0.7843137254901961,0.7843137254901961}{-0.317} & \cbox{0.966320645905421,0.8741253364090734,0.8741253364090734}{-0.226} & \cbox{1.0,0.7843137254901961,0.7843137254901961}{oom} & \cbox{1.0,0.7843137254901961,0.7843137254901961}{oom} \\
&& time &  \cbox{0.9423298731257209,0.9381007304882737,0.9381007304882737}{0.21s} & \cbox{0.9907727797001153,0.8089196462898884,0.8089196462898884}{1.38s} & \cbox{1.0,0.7843137254901961,0.7843137254901961}{5.83s} & \cbox{0.7843137254901961,0.7843137254901961,1.0}{0.02s} & \cbox{0.9221068819684737,0.9221068819684737,0.9483275663206459}{0.15s} & \cbox{0.8458285274894272,0.8458285274894272,0.9769319492502884}{0.05s} & \cbox{0.9741637831603229,0.8532103037293348,0.8532103037293348}{0.72s} & \cbox{0.9497116493656286,0.9184159938485198,0.9184159938485198}{0.28s} & \cbox{1.0,0.7843137254901961,0.7843137254901961}{4.05s} & \cbox{1.0,0.7843137254901961,0.7843137254901961}{oom} & \cbox{1.0,0.7843137254901961,0.7843137254901961}{oom} \\\mymidrulegray
\tworow{ B323 } & \tworow{ 191916 } & value & \cbox{0.7843137254901961,0.7843137254901961,1.0}{0.000} & \cbox{1.0,0.7843137254901961,0.7843137254901961}{-0.375} & \cbox{1.0,0.7843137254901961,0.7843137254901961}{-0.375} & \cbox{0.7843137254901961,0.7843137254901961,1.0}{0.000} & \cbox{0.7843137254901961,0.7843137254901961,1.0}{0.000} & \cbox{1.0,0.7843137254901961,0.7843137254901961}{-0.375} & \cbox{0.9607843137254902,0.8888888888888888,0.8888888888888888}{-0.250} & \cbox{1.0,0.7843137254901961,0.7843137254901961}{-0.375} & \cbox{1.0,0.7843137254901961,0.7843137254901961}{oom} & \cbox{1.0,0.7843137254901961,0.7843137254901961}{oom} & \cbox{1.0,0.7843137254901961,0.7843137254901961}{oom} \\
&& time &  \cbox{0.9319492502883506,0.9319492502883506,0.9446366782006921}{2.82s} & \cbox{1.0,0.7843137254901961,0.7843137254901961}{32.84s} & \cbox{1.0,0.7843137254901961,0.7843137254901961}{1m 55s} & \cbox{0.7843137254901961,0.7843137254901961,1.0}{0.32s} & \cbox{0.9221068819684737,0.9221068819684737,0.9483275663206459}{2.42s} & \cbox{0.930718954248366,0.930718954248366,0.9450980392156862}{2.77s} & \cbox{0.9833910034602076,0.8286043829296424,0.8286043829296424}{16.50s} & \cbox{0.9912341407151095,0.8076893502499038,0.8076893502499038}{22.59s} & \cbox{1.0,0.7843137254901961,0.7843137254901961}{oom} & \cbox{1.0,0.7843137254901961,0.7843137254901961}{oom} & \cbox{1.0,0.7843137254901961,0.7843137254901961}{oom} \\\mymidrulegray
\tworow{ S122 } & \tworow{ 396 } & value & \cbox{0.9695501730103806,0.8655132641291811,0.8655132641291811}{13.636} & \cbox{0.9792387543252595,0.839677047289504,0.839677047289504}{9.565} & \cbox{0.9806228373702423,0.8359861591695502,0.8359861591695502}{9.078} & \cbox{0.7843137254901961,0.7843137254901961,1.0}{50.000} & \cbox{0.7843137254901961,0.7843137254901961,1.0}{50.000} & \cbox{0.9783160322952711,0.8421376393694733,0.8421376393694733}{10.000} & \cbox{0.8347558631295655,0.8347558631295655,0.9810841983852364}{42.000} & \cbox{0.9783160322952711,0.8421376393694733,0.8421376393694733}{10.000} & \cbox{0.8347558631295655,0.8347558631295655,0.9810841983852364}{42.000} & \cbox{1.0,0.7843137254901961,0.7843137254901961}{0.820} & \cbox{0.8347558631295655,0.8347558631295655,0.9810841983852364}{42.000} \\
&& time &  \cbox{0.7843137254901961,0.7843137254901961,1.0}{0.01s} & \cbox{0.8310649750096116,0.8310649750096116,0.9824682814302191}{0.02s} & \cbox{0.8790465205690119,0.8790465205690119,0.9644752018454441}{0.04s} & \cbox{0.7843137254901961,0.7843137254901961,1.0}{0.01s} & \cbox{0.7843137254901961,0.7843137254901961,1.0}{0.01s} & \cbox{0.8310649750096116,0.8310649750096116,0.9824682814302191}{0.02s} & \cbox{0.9437139561707035,0.9344098423683198,0.9344098423683198}{0.11s} & \cbox{0.9257977700884275,0.9257977700884275,0.9469434832756632}{0.08s} & \cbox{0.9589388696655132,0.8938100730488273,0.8938100730488273}{0.20s} & \cbox{0.995847750865052,0.7953863898500577,0.7953863898500577}{0.85s} & \cbox{1.0,0.7843137254901961,0.7843137254901961}{1.74s} \\\mymidrulegray
\tworow{ S123 } & \tworow{ 2376 } & value & \cbox{0.9695501730103806,0.8655132641291811,0.8655132641291811}{13.636} & \cbox{0.9783160322952711,0.8421376393694733,0.8421376393694733}{10.000} & \cbox{0.9783160322952711,0.8421376393694733,0.8421376393694733}{10.000} & \cbox{0.7843137254901961,0.7843137254901961,1.0}{50.000} & \cbox{0.7843137254901961,0.7843137254901961,1.0}{50.000} & \cbox{0.9783160322952711,0.8421376393694733,0.8421376393694733}{10.000} & \cbox{0.8347558631295655,0.8347558631295655,0.9810841983852364}{42.000} & \cbox{0.9783160322952711,0.8421376393694733,0.8421376393694733}{10.000} & \cbox{0.8347558631295655,0.8347558631295655,0.9810841983852364}{42.000} & \cbox{1.0,0.7843137254901961,0.7843137254901961}{0.820} & \cbox{0.8347558631295655,0.8347558631295655,0.9810841983852364}{42.000} \\
&& time &  \cbox{0.8027681660899654,0.8027681660899654,0.9930795847750865}{0.04s} & \cbox{0.9233371780084583,0.9233371780084583,0.9478662053056517}{0.23s} & \cbox{0.9607843137254902,0.8888888888888888,0.8888888888888888}{0.65s} & \cbox{0.7843137254901961,0.7843137254901961,1.0}{0.03s} & \cbox{0.8421376393694733,0.8421376393694733,0.9783160322952711}{0.07s} & \cbox{0.8790465205690119,0.8790465205690119,0.9644752018454441}{0.12s} & \cbox{0.9282583621683967,0.9282583621683967,0.9460207612456747}{0.25s} & \cbox{0.9520184544405997,0.9122645136485967,0.9122645136485967}{0.46s} & \cbox{0.9732410611303345,0.8556708958093041,0.8556708958093041}{1.04s} & \cbox{1.0,0.7843137254901961,0.7843137254901961}{1m 13s} & \cbox{1.0,0.7843137254901961,0.7843137254901961}{1m 49s} \\\mymidrulegray
\tworow{ S133 } & \tworow{ 5632 } & value & \cbox{0.9584775086505191,0.895040369088812,0.895040369088812}{18.182} & \cbox{0.966320645905421,0.8741253364090734,0.8741253364090734}{15.000} & \cbox{0.966320645905421,0.8741253364090734,0.8741253364090734}{15.000} & \cbox{0.7843137254901961,0.7843137254901961,1.0}{50.000} & \cbox{0.7843137254901961,0.7843137254901961,1.0}{50.000} & \cbox{0.966320645905421,0.8741253364090734,0.8741253364090734}{15.000} & \cbox{0.8286043829296424,0.8286043829296424,0.9833910034602076}{43.000} & \cbox{0.966320645905421,0.8741253364090734,0.8741253364090734}{15.000} & \cbox{0.8286043829296424,0.8286043829296424,0.9833910034602076}{43.000} & \cbox{1.0,0.7843137254901961,0.7843137254901961}{0.820} & \cbox{1.0,0.7843137254901961,0.7843137254901961}{oom} \\
&& time &  \cbox{0.7843137254901961,0.7843137254901961,1.0}{0.04s} & \cbox{0.9750865051903114,0.8507497116493656,0.8507497116493656}{1.51s} & \cbox{0.9875432525951557,0.8175317185697809,0.8175317185697809}{2.46s} & \cbox{0.8113802383698577,0.8113802383698577,0.9898500576701269}{0.06s} & \cbox{0.859361783929258,0.859361783929258,0.9718569780853518}{0.12s} & \cbox{0.9233371780084583,0.9233371780084583,0.9478662053056517}{0.31s} & \cbox{0.9534025374855825,0.9085736255286428,0.9085736255286428}{0.65s} & \cbox{0.9792387543252595,0.839677047289504,0.839677047289504}{1.78s} & \cbox{0.9916955017301038,0.8064590542099193,0.8064590542099193}{2.89s} & \cbox{1.0,0.7843137254901961,0.7843137254901961}{17m 7s} & \cbox{1.0,0.7843137254901961,0.7843137254901961}{oom} \\\mymidrulegray
\tworow{ RS12 } & \tworow{ 400 } & value & \cbox{1.0,0.7843137254901961,0.7843137254901961}{6.010} & \cbox{1.0,0.7843137254901961,0.7843137254901961}{6.010} & \cbox{1.0,0.7843137254901961,0.7843137254901961}{6.010} & \cbox{0.7843137254901961,0.7843137254901961,1.0}{6.173} & \cbox{0.7843137254901961,0.7843137254901961,1.0}{6.173} & \cbox{0.7843137254901961,0.7843137254901961,1.0}{6.173} & \cbox{0.7843137254901961,0.7843137254901961,1.0}{6.173} & \cbox{0.7843137254901961,0.7843137254901961,1.0}{6.173} & \cbox{0.7843137254901961,0.7843137254901961,1.0}{6.173} & \cbox{0.7843137254901961,0.7843137254901961,1.0}{6.173} & \cbox{0.7843137254901961,0.7843137254901961,1.0}{6.173} \\
&& time &  \cbox{0.8310649750096116,0.8310649750096116,0.9824682814302191}{0.02s} & \cbox{0.7843137254901961,0.7843137254901961,1.0}{0.01s} & \cbox{0.7843137254901961,0.7843137254901961,1.0}{0.01s} & \cbox{0.7843137254901961,0.7843137254901961,1.0}{0.00s} & \cbox{0.7843137254901961,0.7843137254901961,1.0}{0.01s} & \cbox{0.7843137254901961,0.7843137254901961,1.0}{0.00s} & \cbox{0.7843137254901961,0.7843137254901961,1.0}{0.01s} & \cbox{0.7843137254901961,0.7843137254901961,1.0}{0.01s} & \cbox{0.8790465205690119,0.8790465205690119,0.9644752018454441}{0.04s} & \cbox{0.9976931949250288,0.7904652056901191,0.7904652056901191}{0.91s} & \cbox{1.0,0.7843137254901961,0.7843137254901961}{1.77s} \\\mymidrulegray
\tworow{ RS13 } & \tworow{ 4356 } & value & \cbox{0.9856978085351787,0.8224529027297194,0.8224529027297194}{9.398} & \cbox{0.9916955017301038,0.8064590542099193,0.8064590542099193}{9.385} & \cbox{1.0,0.7843137254901961,0.7843137254901961}{9.367} & \cbox{0.7843137254901961,0.7843137254901961,1.0}{9.622} & \cbox{0.7843137254901961,0.7843137254901961,1.0}{9.622} & \cbox{0.7843137254901961,0.7843137254901961,1.0}{9.622} & \cbox{0.7843137254901961,0.7843137254901961,1.0}{9.622} & \cbox{0.7843137254901961,0.7843137254901961,1.0}{9.622} & \cbox{0.7843137254901961,0.7843137254901961,1.0}{9.622} & \cbox{0.8212226066897347,0.8212226066897347,0.986159169550173}{9.592} & \cbox{0.8212226066897347,0.8212226066897347,0.986159169550173}{9.592} \\
&& time &  \cbox{0.9986159169550173,0.78800461361015,0.78800461361015}{2.82s} & \cbox{1.0,0.7843137254901961,0.7843137254901961}{1m 28s} & \cbox{1.0,0.7843137254901961,0.7843137254901961}{12m 31s} & \cbox{0.7843137254901961,0.7843137254901961,1.0}{0.03s} & \cbox{0.8728950403690888,0.8728950403690888,0.9667820069204152}{0.11s} & \cbox{0.8421376393694733,0.8421376393694733,0.9783160322952711}{0.07s} & \cbox{0.9098039215686274,0.9098039215686274,0.9529411764705882}{0.19s} & \cbox{0.8987312572087658,0.8987312572087658,0.9570934256055363}{0.16s} & \cbox{0.9566320645905421,0.8999615532487505,0.8999615532487505}{0.55s} & \cbox{1.0,0.7843137254901961,0.7843137254901961}{3m 9s} & \cbox{1.0,0.7843137254901961,0.7843137254901961}{6m 42s} \\\mymidrulegray
\tworow{ RS14 } & \tworow{ 229888 } & value & \cbox{1.0,0.7843137254901961,0.7843137254901961}{oom} & \cbox{1.0,0.7843137254901961,0.7843137254901961}{oom} & \cbox{1.0,0.7843137254901961,0.7843137254901961}{oom} & \cbox{0.7843137254901961,0.7843137254901961,1.0}{10.500} & \cbox{0.7843137254901961,0.7843137254901961,1.0}{10.500} & \cbox{0.7843137254901961,0.7843137254901961,1.0}{10.500} & \cbox{0.7843137254901961,0.7843137254901961,1.0}{10.500} & \cbox{0.7843137254901961,0.7843137254901961,1.0}{10.500} & \cbox{0.7843137254901961,0.7843137254901961,1.0}{10.500} & \cbox{1.0,0.7843137254901961,0.7843137254901961}{oom} & \cbox{1.0,0.7843137254901961,0.7843137254901961}{oom} \\
&& time &  \cbox{1.0,0.7843137254901961,0.7843137254901961}{oom} & \cbox{1.0,0.7843137254901961,0.7843137254901961}{oom} & \cbox{1.0,0.7843137254901961,0.7843137254901961}{oom} & \cbox{0.7843137254901961,0.7843137254901961,1.0}{0.20s} & \cbox{0.9171856978085352,0.9171856978085352,0.9501730103806229}{1.41s} & \cbox{0.8655132641291811,0.8655132641291811,0.9695501730103806}{0.66s} & \cbox{0.9589388696655132,0.8938100730488273,0.8938100730488273}{3.97s} & \cbox{0.9450980392156862,0.930718954248366,0.930718954248366}{2.34s} & \cbox{0.9870818915801615,0.8187620146097655,0.8187620146097655}{12.04s} & \cbox{1.0,0.7843137254901961,0.7843137254901961}{oom} & \cbox{1.0,0.7843137254901961,0.7843137254901961}{oom} \\\mymidrulegray
\tworow{ RS22 } & \tworow{ 484 } & value & \cbox{0.9967704728950404,0.7929257977700884,0.7929257977700884}{7.188} & \cbox{1.0,0.7843137254901961,0.7843137254901961}{7.176} & \cbox{1.0,0.7843137254901961,0.7843137254901961}{7.176} & \cbox{0.7843137254901961,0.7843137254901961,1.0}{7.594} & \cbox{0.7843137254901961,0.7843137254901961,1.0}{7.594} & \cbox{0.7843137254901961,0.7843137254901961,1.0}{7.594} & \cbox{0.7843137254901961,0.7843137254901961,1.0}{7.594} & \cbox{0.7843137254901961,0.7843137254901961,1.0}{7.594} & \cbox{0.7843137254901961,0.7843137254901961,1.0}{7.594} & \cbox{0.7843137254901961,0.7843137254901961,1.0}{7.594} & \cbox{0.7843137254901961,0.7843137254901961,1.0}{7.594} \\
&& time &  \cbox{0.9589388696655132,0.8938100730488273,0.8938100730488273}{0.20s} & \cbox{0.9589388696655132,0.8938100730488273,0.8938100730488273}{0.20s} & \cbox{0.9534025374855825,0.9085736255286428,0.9085736255286428}{0.16s} & \cbox{0.7843137254901961,0.7843137254901961,1.0}{0.00s} & \cbox{0.7843137254901961,0.7843137254901961,1.0}{0.01s} & \cbox{0.7843137254901961,0.7843137254901961,1.0}{0.01s} & \cbox{0.8310649750096116,0.8310649750096116,0.9824682814302191}{0.02s} & \cbox{0.7843137254901961,0.7843137254901961,1.0}{0.01s} & \cbox{0.8790465205690119,0.8790465205690119,0.9644752018454441}{0.04s} & \cbox{0.9976931949250288,0.7904652056901191,0.7904652056901191}{0.90s} & \cbox{1.0,0.7843137254901961,0.7843137254901961}{1.80s} \\\mymidrulegray
\tworow{ RS23 } & \tworow{ 4096 } & value & \cbox{0.972318339100346,0.8581314878892734,0.8581314878892734}{10.961} & \cbox{0.9953863898500577,0.7966166858900423,0.7966166858900423}{10.820} & \cbox{1.0,0.7843137254901961,0.7843137254901961}{10.791} & \cbox{0.7843137254901961,0.7843137254901961,1.0}{11.516} & \cbox{0.7843137254901961,0.7843137254901961,1.0}{11.516} & \cbox{0.7843137254901961,0.7843137254901961,1.0}{11.513} & \cbox{0.7843137254901961,0.7843137254901961,1.0}{11.513} & \cbox{0.7966166858900423,0.7966166858900423,0.9953863898500577}{11.485} & \cbox{0.7966166858900423,0.7966166858900423,0.9953863898500577}{11.485} & \cbox{0.8064590542099193,0.8064590542099193,0.9916955017301038}{11.464} & \cbox{0.8064590542099193,0.8064590542099193,0.9916955017301038}{11.464} \\
&& time &  \cbox{1.0,0.7843137254901961,0.7843137254901961}{3.12s} & \cbox{1.0,0.7843137254901961,0.7843137254901961}{56m 31s} & \cbox{1.0,0.7843137254901961,0.7843137254901961}{6m 35s} & \cbox{0.7843137254901961,0.7843137254901961,1.0}{0.03s} & \cbox{0.8655132641291811,0.8655132641291811,0.9695501730103806}{0.10s} & \cbox{0.8310649750096116,0.8310649750096116,0.9824682814302191}{0.06s} & \cbox{0.9061130334486736,0.9061130334486736,0.9543252595155709}{0.18s} & \cbox{0.9098039215686274,0.9098039215686274,0.9529411764705882}{0.19s} & \cbox{0.960322952710496,0.8901191849288735,0.8901191849288735}{0.63s} & \cbox{1.0,0.7843137254901961,0.7843137254901961}{7m 43s} & \cbox{1.0,0.7843137254901961,0.7843137254901961}{12m 1s} \\\mymidrulegray

\end{tabular}
}
\label{ta:experiments}
\end{table}

\newcommand{\payoffspace}[3]{%
\begin{tikzpicture}
  \node at (0,0) {\includegraphics[scale=.6]{plots/#1#2#3_payoff_space.pdf}};
  \ifthenelse{#2=3}{
    \node[anchor=north west] at (-2,2) {\scalebox{.8}{$^{#2}$#1#3}};
  }{
    \node[anchor=south] at (.5,2.05) {\scalebox{.8}{#1#3}};
  }
\end{tikzpicture}
}%

\begin{figure}[!tb]\centering
\payoffspace {battleship}2{222}
\payoffspace {sheriff}2{122}
\payoffspace {rideshare}2{22}

\tikzstyle{efce}=[fill=orange!15!white,draw=orange,densely dotted]
\tikzstyle{efcce}=[fill=none,draw=orange,thin,densely dashed]
\tikzstyle{nfcce}=[fill=none,draw=orange,thin]
\tikzstyle{nfccert}=[fill=none,draw=blue,thick]
\tikzstyle{ccert}=[fill=none,draw=blue,thick,densely dashed]
\tikzstyle{cert}=[fill=blue!15!white,draw=blue,thick,densely dotted]
\tikzstyle{comm}=[fill=green!15!white,fill opacity=1,draw=green,thick,densely dotted]

    \tikzstyle{lbl}=[inner sep=0mm]

    \begin{tikzpicture}

        \begin{scope}
            \filldraw[nfcce] (0, 0) rectangle +(.7,.25);
            \node[lbl,anchor=south west] at (.85, 0) {\small NFCCE};
        \end{scope}
        \begin{scope}[xshift=3.4cm]
            \filldraw[efcce] (0, 0) rectangle +(.7,.25);
            \node[lbl,anchor=south west] at (.85, 0) {\small EFCCE};
        \end{scope}
        \begin{scope}[xshift=6.8cm]
            \filldraw[efce] (0, 0) rectangle +(.7,.25);
            \node[lbl,anchor=south west] at (.85, 0) {\small EFCE};
        \end{scope}
        \begin{scope}[xshift=10.2cm]
            \node[lbl,anchor=south west] at (.85, 0) {\small \phantom{({\smaller $\bigstar$}) Communication}};
        \end{scope}
    \end{tikzpicture}
    \\
    \begin{tikzpicture}
        \begin{scope}[xshift=0cm]
            \filldraw[nfccert] (0, 0) rectangle +(.7,.25);
            \node[lbl,anchor=south west] at (.85, 0) {\small NF coarse full-cert};
        \end{scope}
        \begin{scope}[xshift=3.4cm]
            \filldraw[ccert] (0, 0) rectangle +(.7,.25);
            \node[lbl,anchor=south west] at (.85, 0) {\small  Coarse full-cert};
        \end{scope}
        \begin{scope}[xshift=6.8cm]
            \filldraw[cert] (0, 0) rectangle +(.7,.25);
            \node[lbl,anchor=south west] at (.85, 0) {\small  Full-certification};
        \end{scope}
        \begin{scope}[xshift=10.2cm]
            \filldraw[comm] (0, 0) rectangle +(.7,.25);
            \node[lbl,anchor=south west] at (.85, 0) {\small (\textcolor{green}{\smaller $\bigstar$}) Communication};
        \end{scope}
    \end{tikzpicture}

\caption{Payoff spaces for various games and notions of equilibrium. The symbol \textcolor{green}{\smaller $\bigstar$} indicates that the set of communication equilibrium payoffs for that game is (at least, modulo numerical precision) that single point. In the battleship instance, many of the notions overlap. }\label{fi:payoff-spaces}
\end{figure}
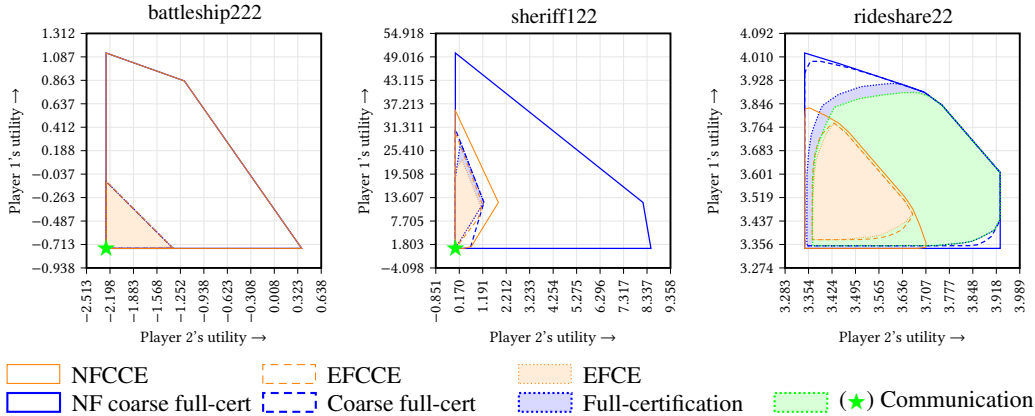
We ran our algorithm for communication and full-certification equilibria on various two-player games, and compared the results to those given by notions of optimal correlation in games. The games used in the experiments are given in \Cref{se:experiment-games}. 
All experiments were allocated four CPU cores and 64 GB of RAM. Linear programs were solved with Gurobi 9.5. When payments are used, the allowable payment range is $[0, M]$ where $M$ is the reward range of the game.
Experimental results can be found in \Cref{ta:experiments}. 

In the {\em battleship} and {\em sheriff} instances, there is not a significant difference in performance between finding full-certification equilibria and finding optimal correlated equilibria in terms of performance---this is because, unlike in the general case, optimal correlated equilibria in two-player games without chance can be found in polynomial time~\cite{Stengel08:Extensive} anyway. In the {\em ridesharing} instances, computing optimal correlated equilibria is much more computationally intensive because the game contains non-public chance actions. Computing optimal full-certification equilibria is comparably easy, and this difference is clearly seen in the timing results. 

Finding optimal {\em communication} equilibria is much more intensive than finding optimal full-certification equilibria, owing to the quadratic size of the augmented game for communication equilibria. This often causes communication equilibria to be the {\em hardest} of the notions to compute in practice, despite optimal correlation being NP-hard.

In \Cref{fi:payoff-spaces}, we have plotted the payoff spaces of some representative instances. The plots show how the polytopes of communication and full-certification equilibria behave relative to correlated equilibria. In the {\em battleship} and {\em sheriff} instances, the space of communication equilibrium payoffs is a single point, which implies that the space of NFCE (and hence Nash) equilibrium payoffs is also that single point. Unfortunately, that point is the Pareto-least-optimal point in the space of EFCEs. In the {\em ridesharing} instances, communication allows higher payoffs. This is because the mediator is allowed to ``leak'' information between players.

\section{Conclusions and future research}
We have shown that optimal communication and certification equilibria in extensive-form games can be computed via linear programs of polynomial size, or almost-linear size in the full-certification case. We have used our machinery to derive an entire family of equilibrium concepts which we hope to be of use in the future. 

Possible future directions include the following.
\begin{enumerate}
    \item Are there efficient {\em online learning dynamics}, in any reasonable sense of that term, that converge to certification or communication equilibrium?
    \item Is there a better-than-quadratic-size linear program for communication equilibria?
    \item Is it possible to extend our augmented game construction to also cover {\em normal-form} correlated equilibria while maintaining efficiency?
    \item Investigate further the comparison between communication and correlation in games. For example, when and why do communication equilibria achieve higher social welfare than extensive-form correlated equilibria?
\end{enumerate}

\section*{Acknowledgements}
This material is based on work supported by the National Science Foundation under grants \mbox{IIS-1901403} and \mbox{CCF-1733556}, and the ARO under award W911NF2010081. 

\newpage
\bibliographystyle{plainnat}
\bibliography{dairefs}

\section*{Checklist}

\begin{enumerate}

\item For all authors...
\begin{enumerate}
  \item Do the main claims made in the abstract and introduction accurately reflect the paper's contributions and scope?
    \answerYes{}
  \item Did you describe the limitations of your work?
    \answerYes{}
  \item Did you discuss any potential negative societal impacts of your work?
    \answerNA{}
  \item Have you read the ethics review guidelines and ensured that your paper conforms to them?
    \answerYes{}
\end{enumerate}

\item If you are including theoretical results...
\begin{enumerate}
  \item Did you state the full set of assumptions of all theoretical results?
    \answerYes{}
        \item Did you include complete proofs of all theoretical results?
    \answerYes{in appendix}
\end{enumerate}

\item If you ran experiments...
\begin{enumerate}
  \item Did you include the code, data, and instructions needed to reproduce the main experimental results (either in the supplemental material or as a URL)?
    \answerNo{}
  \item Did you specify all the training details (e.g., data splits, hyperparameters, how they were chosen)?
    \answerYes{}
        \item Did you report error bars (e.g., with respect to the random seed after running experiments multiple times)?
    \answerNA{Experiments were deterministic.}
        \item Did you include the total amount of compute and the type of resources used (e.g., type of GPUs, internal cluster, or cloud provider)?
    \answerYes{}
\end{enumerate}

\item If you are using existing assets (e.g., code, data, models) or curating/releasing new assets...
\begin{enumerate}
  \item If your work uses existing assets, did you cite the creators?
    \answerNA{}
  \item Did you mention the license of the assets?
    \answerNA{}
  \item Did you include any new assets either in the supplemental material or as a URL?
    \answerNA{}
  \item Did you discuss whether and how consent was obtained from people whose data you're using/curating?
    \answerNA{}
  \item Did you discuss whether the data you are using/curating contains personally identifiable information or offensive content?
    \answerNA{}
\end{enumerate}

\item If you used crowdsourcing or conducted research with human subjects...
\begin{enumerate}
  \item Did you include the full text of instructions given to participants and screenshots, if applicable?
    \answerNA{}
  \item Did you describe any potential participant risks, with links to Institutional Review Board (IRB) approvals, if applicable?
    \answerNA{}
  \item Did you include the estimated hourly wage paid to participants and the total amount spent on participant compensation?
    \answerNA{}
\end{enumerate}

\end{enumerate}

\newpage
\appendix
\section{Omitted proofs}

\subsection{\Cref{pr:rev}}

\begin{proof}
We follow the usual structure of revelation principle proofs. Given an \S-certification equilibrium, consider augmenting the mediator $d$ by devices $d^i$, one per player, that acts as follows. $d^i$ internally keeps track of the current sequence $\sigma_i$ of player $i$. Whenever it is player $i$'s turn to act, $d^i$ expects to be told an information set $I'$ with $\sigma(I') = Ia$. Then $d^i$ samples what message $s$ player $i$ would have honestly sent to the mediator at infoset $I'$, and forwards this to $d$. When $d$ replies with a message $s'$, $d^i$ samples the action $a'$ the player would have played, sends that to the player, and updates her internal state to $\sigma_i := I'a'$. If $d^i$ ever receives an invalid message (i.e., anything except an infoset $I'$ with $\sigma(I') = Ia$), it resorts to always sending $\bot$ to both $d$ and player $i$ for the remainder of the game.

To see that this results in an equilibrium, note that player $i$ can always simulate $d^i$ when playing with the original mediator $d$---therefore, any deviation she can perform against the direct equilibrium can also be performed with $d$ (This is where NRC is used: since the acting player may in reality have lied about her information, without the NRC, it could be the case that the simulated message $s$ would not be legal for player $i$ to send).
\end{proof}
\subsection{\Cref{pr:rev2}}

\begin{proof}
Consider a mediator for the full game that acts according to $\hat x_\mediator$. If the player sends messages that are not in $\hat\Gamma$, then the mediator acts as if the player sent $\bot$. If the mediator reaches a state in which at least two players have provably deviated (i.e., the mediator has reached a state not in $\hat\Gamma$), then the mediator acts arbitrarily. This mediator strategy, along with the players' direct strategies, forms a strategy profile. The only difference between $\hat\Gamma$ and the true communication protocol induced by \S is, in $\hat\Gamma$, players are not able to send certain messages. But, in any case, those messages are ones that are never sent by an honest player; therefore, the mediator will always act as if the player sent $\bot$ if it receives such a message. Therefore, the player gains nothing by having extra messages that she can send. This completes the proof. 
\end{proof}

\section{Full linear program}\label{se:lp-full}
In this section, we present the full LP formulation implied by the discussion in \Cref{se:proof-main}.

Dropping the hats for notational cleanliness, and letting $\mc X_j = \{ \vec x_j  : \vec F_j \vec x_j = \vec f_j, \vec x_j \ge 0 \}$ be the realization-form representation of player $j$'s decision space, the program \eqref{eq:program} has the form
\begin{align}
    \max_{{\vec x}_\mediator \in {\mc X}_\mediator} \vec c^\top {\vec x} \qq{s.t.} \max_{\vec x_j : \vec F_j \vec x_j = \vec f_j, \vec x_j \ge 0} \vec x^\top_\mediator \vec A_j \vec x_j \le 0.~~\forall j \in [n]
\end{align}
Dualizing the inner maximizations, we have the linear program
\begin{align}
    \max~~&\vec c^\top {\vec x}_\mediator \\
    \qq{s.t.}&
    \vec F_j^\top \vec v_j \ge \vec A_j^\top \vec x_\mediator,~~\vec f_j^\top \vec v_j \le 0~~\forall j \in [n]
    \\& \vec x_\mediator \in {\mc X}_\mediator
\end{align}
where $\vec v_j$ are dual variables. Intuitively, the dual variables represent {\em best-response values}: the vector $\vec v_j$ will be indexed by information sets $I$ for player $j$, and $v_j[I]$ will be the best-response value for player $j$ at infoset $I$ (assuming all other players are direct). This sort of dualization of an inner optimization problem to create a linear program is fairly standard---see \citet{Koller94:Fast} for an application to two-player zero-sum games and \citet{Farina19:Correlation} for the same approach applied to the special case of correlation.

\section{Example of game in which full-certification equilibria dominate NFCCEs}\label{se:counterexample}

\begin{figure}[!htb]
\newcommand{\util}[2]{\sf\footnotesize {\color{p1color} #1}, {\color{p2color} #2}} 
\centering
\tikzset{
    every path/.style={-stealth},
    infoset1/.style={-, dotted, ultra thick, color=p1color},
    infoset2/.style={-, dotted, ultra thick, color=p2color},
    terminal/.style={font=\sf, draw=none},
    s1/.style={}, %
    s2/.style={}, %
}
\forestset{
    p1/.style={regular
        polygon, regular polygon
        sides=3, inner sep=2pt, fill=p1color, draw=none},
    p2/.style={p1, shape border rotate=60, fill=p2color},
    parent/.style={no edge,tikz={\draw (#1.south) -- (!.north); }},
    el/.style={edge label={node[midway, fill=white, inner sep=1pt] {\sf\tiny #1}}}
}
\begin{forest}
for tree={parent anchor=south, child anchor=north}
[,
[,p1
	[\util{0}{1},terminal]
	[,p2
		[,p1,name=3
			[\util{1}{--1},terminal]
			[\util{--1}{1},terminal]
		]
		[,p1,name=4
			[\util{--1}{1},terminal]
			[\util{1}{--1},terminal]
		]
	]
]
[,p2
	[\util{1}{0},terminal]
	[,p1
		[,p2,name=1
			[\util{1}{--1},terminal]
			[\util{--1}{1},terminal]
		]
		[,p2,name=2
			[\util{--1}{1},terminal]
			[\util{1}{--1},terminal]
		]
	]
]
[,p1,name=5,el=H
	[\util{1}{1},terminal]
	[\util{0}{0},terminal]
]
[,p1,name=6,el=T
	[\util{0}{0},terminal]
	[\util{1}{1},terminal]
]
]
\draw[infoset2] (1) -- (2);
\draw[infoset1] (3) -- (4);
\draw[infoset1] (5) -- (6);
\end{forest}
\caption{The counterexample in \Cref{se:counterexample}.
At terminal nodes, \pone's utility is listed first. The root node is a chance node, and the distribution at it is uniform random.}
	\label{fi:pareto-dominance}
\end{figure}
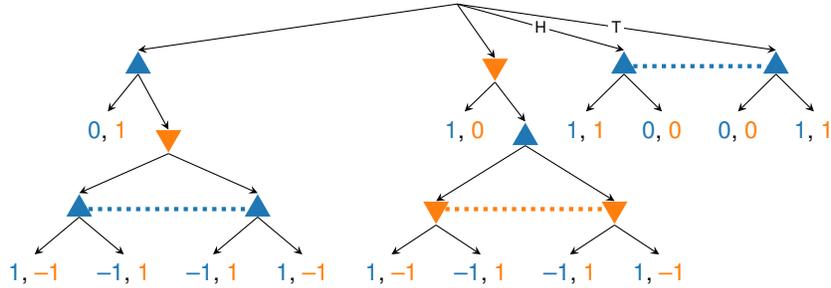

We now give an example that demonstrates the difference between NFCCEs (where the mediator has imperfect recall) and the perfect-recall notions. In particular, we will show a game where there is an (extensive-form) persuasion strategy for the mediator that leads to outcomes Pareto-dominating every NFCCE of $\Gamma$. Since persuasion can be recovered from full certification (or even communication) by adding a dummy player who communicates information to the mediator and has no incentives (thus is always direct), this also shows that communication equilibria can Pareto-dominate NFCCEs.

Consider the game whose tree is depicted in \Cref{fi:pareto-dominance}. At the root, chance selects a subgame $s \in \{\pone, \ptwo, {\sf H}, {\sf T}\}$. If chance selects $s \in \{\pone, \ptwo\}$,  then player $s$ is offered a chance to exit the game; if she does, she scores $0$ and the other player scores $1$. If she does not exit, the two players play matching pennies with the {\em other} player (not $s$) playing first (this will be relevant when we take perfect-recall refinements). The winner scores $1$ point and the loser $-1$. If chance selects $s \in \{{\sf H}, {\sf T}\}$, then \pone attempts to guess which one was selected. If successful, both players score $1$ point; otherwise, both scores $0$. The following properties can be directly verified:
\begin{itemize}
\item For every NFCCE in both $\Gamma$ and its perfect-information refinement, both \pone and \ptwo have value exactly $1/2$.
\item The following mediator pure strategy is persuasive in $\Gamma$, where both players achieve value $3/4$: both players exit when possible, and if $s \in \{ {\sf H}, {\sf T} \}$, the mediator informs \pone which is the case. If the matching pennies subgame is entered, the mediator recommends that each player play independently and uniformly at random.
\end{itemize}
Intuitively, this outcome exists because the mediator has the power to selectively tell players information when it is to their benefit (\eg in the case $s \in \{ {\sf H}, {\sf T}\}$, where the mediator can inform \pone of chance's choice to allow \pone to score a point) and withhold information to incentivize cooperation (\eg, in the case $s = \pone$, where the mediator can withhold \ptwo's matching pennies bit from \pone to prevent \pone from deviating).

\section{Descriptions of games in experiments}\label{se:experiment-games}

\begin{itemize}
    \item {\bf B} is a two-player Battleship game~\cite{Farina19:Correlation} with a grid of size $h \times w$ and $r$ rounds, made into a nonzero-sum game by having each player value their own ships greater than their opponents’. The three numbers afterward are, in order, the length, width, and number of shots per side. Each player has a single ship of unit size.
\item {\bf S} is a simplified Sheriff of Nottingham game~\cite{Farina19:Correlation}, a small game modeling a negotiation between two players. The first two numbers roughly correspond to how much power each player has in negotiation; the final number is the number of rounds of negotiation.
\item {\bf RS} is a small ridesharing game~\cite{Zhang22:Optimal}, played on a graph, in which the players walk around the graph attempting to serve “customers” who reside at certain nodes. The first number (1 or 2) denotes the specific graph on which the game is played; the second number is the number of steps in the game.
\end{itemize}

\section{Mechanisms with payments}\label{se:payments}

The mediator may be able to take its own actions during the game, separate from the players. In our framework, this can be modelled by introducing an auxiliary player that has no incentives of her own and therefore no reason to disobey the mediator recommendations. One particular use case of such a player is {\em payments}: the auxiliary player can collect payments from or give payments to players, as a means of incentivizing players to perform certain actions.

We can make this concrete as follows. At the end of the game $\hat\Gamma$, chance picks a random player $i^* \in [n]$, reveals it to the mediator. The mediator then selects a payment $p \in \{ L, U \}$ to be given to the player, where $L, U \in \R$ are a minimum or maximum payment (that may be functions of the mediator's information). The player gains utility $np$ and the mediator loses utility $np$. By varying the probability with which the mediator makes the two payments, the mediator can (in expectation) pay any player any amount in the range $[L, U]$. The size of the mediator-augmented game increases by a factor of $O(n)$.

If $U-L$ is at least the reward range of the game, and the mediator always learns whether players have acted directly, then any strategy at all can be enforced in equilibrium simply by giving each player a sufficiently large payoff for cooperating. The ability of the mediator to commit to a strategy is critical when payments are involved: if a mediator could not credibly commit, then it would never make a payment because it has negative incentive to do so. 

{\bf Automated mechanism design.} A consequence of the above analysis is that we recover the Bayes-Nash\footnote{There is no correlation to speak of, because players do not have any actions---they merely report information. Hence, we are able to discuss Bayes-Nash instead of Bayes-correlated equilibria.} randomized automated mechanism design algorithm of \citet{Conitzer02:Complexity,Conitzer04:Self} as a special case, as follows. That paper considers a Bayesian game with $T$ types per player and $n$ players. A type assignment $\vec\theta \in [T]^n$ is sampled from some joint distribution, and each type $\theta_i$ is revealed privately to player $i$ (for {\em ex interim} incentive compatibility) or publicly to all players (for {\em ex post} incentive compatibility). A round of communication ensues, in which each player informs the mediator about her own type (or lies about it). The mediator then chooses an outcome $o \in [O]$, and each player receives a utility that is a function of their own type $\theta_i$ and the outcome $o$. The resulting game has size $T^nO$, and each player has at most $T^n$ sequences, so the overall LP \eqref{eq:program}, after including payments, has size $\poly(O, T^n)$, which is the same result observed by \citet{Conitzer04:Self}.

{\bf Costly messaging.} 
Suppose that players have a strict preference for not revealing information over revealing it, or for sending certain messages over others. In the game $\hat\Gamma$, we can easily express this preference by changing the utilities of player $i$.
In this setting, direct equilibria may not exist in the general case: indeed, consider a single-player game where the player has perfect information and therefore no need to tell the mediator anything. Nonetheless, direct equilibria can still exist in some games if the mediator has some power to persuade the players to reveal their information---that is, if a player's information will help the mediator to give the player a better outcome. If the mediator knows whether players have been direct, there always exist payments large enough to incentivize direct behavior (or, indeed, {\em any} behavior), so payments can also be used to recover equilibrium existence. 

This algorithm does not take into account the possibility that the mediator or players may be better off {\em not knowing} certain information than paying the price of gathering it, which is often the case in preference elicitation~[\eg,~\citenum{Larson05:Mechanism,Sandholm06:Preference}] and has more recently been studied for automated mechanism design~\citet{Zhang21:Automatedb,Kephart15:Complexity,Kephart21:Revelation}; it will only compute the optimal equilibrium for which the mediator gives sufficient incentive for all players to always honestly reveal information. Finding optimal equilibria in settings when the revelation  principle fails is beyond the scope of this paper.

\section{Additional related research}\label{se:related}

Our results are extremely general across settings and applications, most notably mechanism design and Bayesian persuasion (information design). Many special cases of our main algorithm have been discussed in the literature. Here, we give an overview of some of them.

{\bf Automated multi-stage mechanism design.}
In mechanism design, an uninformed mediator (the mechanism) takes actions based on information given to them by player(s) who have information but cannot take actions. In the previous section, we showed how our framework can be used to generalize the automated mechanism design algorithm of \citet{Conitzer02:Complexity}. We will now discuss several other very recent papers on automated mechanism design in dynamic settings, and how our paper relates to them.

\citet{Zhang21:Automated} consider an automated dynamic mechanism design setting in a Markov game with one player. Their main positive result is an LP for the setting of short-horizon MDPs, which can be viewed as a special case of our framework by simply unrolling the (short-horizon) MDP into an extensive-form game.

\citet{Zhang22:Planning} consider an automated mechanism design setting in which the agent's only power is, at each timestep, to {\em quit} the decision process. They study Markov games, which in this setting are significantly involved difficult than extensive-form games. In extensive form, their setting can be formulated as an augmented game very similar to our framework, in which the ``exit'' action is explicitly added into the augmented game and the principal selects the action. 

\citet{Papadimitriou22:Complexity} study a dynamic auction design setting. In their most general setup, there are $k$ players (bidders) with independent valuations for each of $D$ items to be sold in sequence. The agents in their setting know {\em all} their valuations upfront, but only {\em report} their valuations for the item currently being sold. That constraint can be expressed in our framework in the language of $(\mc S, \mc M)$-certification equilibria, by setting $\mc M$ such that the mediator only learns the player's valuation of the current item. As such, that setting is also a special case of ours---in particular, our algorithm matches their positive results,  Theorems~7 and~8, that use a ``dynamic programming LP'' to compute the optimal randomized mechanism.

\citet{Sandholm07:Automated} study multi-stage mechanism design, but they use multiple stages for the purpose of reducing the amount of communication necessary for preference elicitation in what would otherwise be a single-stage setting. Their work is therefore orthogonal to ours. 

{\bf Automated mechanism design with partially-verifiable types.} \citet{Zhang21:Automatedb} give an algorithm for finding the optimal {\em direct} mechanism in a single-agent, single-stage setting with partially-verifiable types. The positive results in their paper focus on the case when all types have the same preferences over outcomes. Our work differs from that one in that we consider extensive-form (multi-stage) settings and a more general setup (in which the players can also take actions and can have arbitrary utility functions). However, we share in common with that paper the fact that we only compute the optimal {\em direct} mechanism, and hence rely on the revelation principle holding for that mechanism to be optimal across communication structures.

\citet{Kephart15:Complexity} analyze mechanism design in a single-stage, single-agent setting with costly reporting and not assuming the revelation principle. Their goal is to, given a social choice function, determine whether that function can be implemented. They analyze a large spectrum of cases, and discuss for each case whether this implementation problem is easy or (NP-)hard. In a follow-up paper~\cite{Kephart21:Revelation}, the same authors carefully investigate when the revelation principle does or does not hold (still in the single-stage, single-agent setting with reporting costs). When the revelation principle does hold, as we have discussed in \Cref{se:payments}, our framework matches, as a special case, the polynomial-time algorithm of \citet{Conitzer02:Complexity,Conitzer04:Self}, and both can be used to compute whether a social choice function is implementable, simply by adding the appropriate linear constraints to the linear program formulation. However, when the revelation principle fails, that approach will fail to find an implementation for any social choice function that cannot be implemented by a direct mechanism.

{\bf Automated multi-stage Bayesian persuasion (information design).}
In Bayesian persuasion, also commonly referred to as information resign~\cite{Kamenica11:Bayesian}, the roles of the mediator and player are reversed compared to automated mechanism design: the mediator (``principal'') has informational advantage, and the player(s) take the actions. The difference between the two settings lies in who has the commitment power: in automated mechanism design, the side with the power to take actions has the commitment power.

\citet{Celli20:Private} study persuasion in extensive-form games with multiple players, in which the principal can only send a single signal at the beginning of the game to the players. They focus on a setting where the mediator can only send a single signal to the players, which is far more restrictive than our setup of persuasion. Their setting of Bayes {\em coarse}-correlated equilibria is equivalent to what we call {\em normal-form coarse imperfect-recall persuasion}, and their results in this setting fall out as special case of our framework (using the correlation DAG of \citet{Zhang22:Optimal} to represent the mediator's decision space). Their discussion of Bayes-{\em correlated} equilibria is not captured by our framework for the same reason that NFCEs are not captured.

\citet{Gan22:Bayesian} study persuasion in Markov games with a single player, in which the principal can send messages to the player at every time step. The extensive-form analogue of their setting is, once again, a special case of our framework. The complexity gap between myopic, advice-myopic, and far-sighted players, discussed in that paper for Markov games, disappears in extensive-form games because extensive form naturally allows history-dependent strategies for both the mediator and the players.

\citet{Wu22:Sequential} study persuasion in Markov games with myopic players. That is, a new player arrives at every timestep. After performing a single action, the player gains a reward dependent on the true state (which, of course, the player may not actually know) and her action, and then leaves the system forever. The authors devise an online learning algorithm that provably has low regret for the sender. Compared to their setting, our setting is significantly more general---allowing multiple non-myopic agents as well as other forms of limited information and communication---but ours is based on linear programming instead of online learning, and works with extensive-form games instead of Markov games.

We are not the first to point out associations between communication equilibria and other settings such as mechanism design and persuasion. In particular, \citet{Bergemann19:Information} discuss in great depth the relationship between communication equilibria, mechanism design, and persuasion in Bayesian games. On top of that paper, our contribution is that we extend their ideas of unification to the more general setting of arbitrary extensive-form games and provide efficient algorithms for all of those cases.

{\bf Preference elicitation from multiple agents.}
There has been significant work on preference elicitation from multiple agents, starting in the context of combinatorial auctions~\citep{Conen01:Preference}. Often the elicitor knows that some preference information from an agent is not needed for optimally allocating (and pricing) given what other agents have already revealed about their preferences. This is the case already in private-values settings. In quasilinear settings, if the elicitor receives enough information from the agents to determine the welfare-maximizing allocation and the Vickrey-Clarke-Groves payments, it is an \textit{ex post} equilibrium for all players to tell the truth~\citep{Conen01:Preference}. This is despite the fact that the elicitor's queries leak information across agents.  Similar leakage also happens in preference elicitation in voting, and there are certain restrictions that can be imposed on the elicitation policy that eliminate any strategic effects that stem from that~\cite{Conitzer02:Vote}. 

Our paradigm is fundamentally different from the above preference elicitation setting, in several ways. First, all our complexity results are at least linear in the size of the game tree, whereas in a combinatorial auction, the auctioneer (mediator) may pick one of exponentially many outcomes. Second, we always allow mixed strategies, which are typically not used in preference elicitation. Finally, in voting settings one often desires an equilibrium in {\em weakly dominant} strategies, which our paradigm also does not cover.

\newpage
\section{Inclusion hierarchy}\label{se:inclusions}

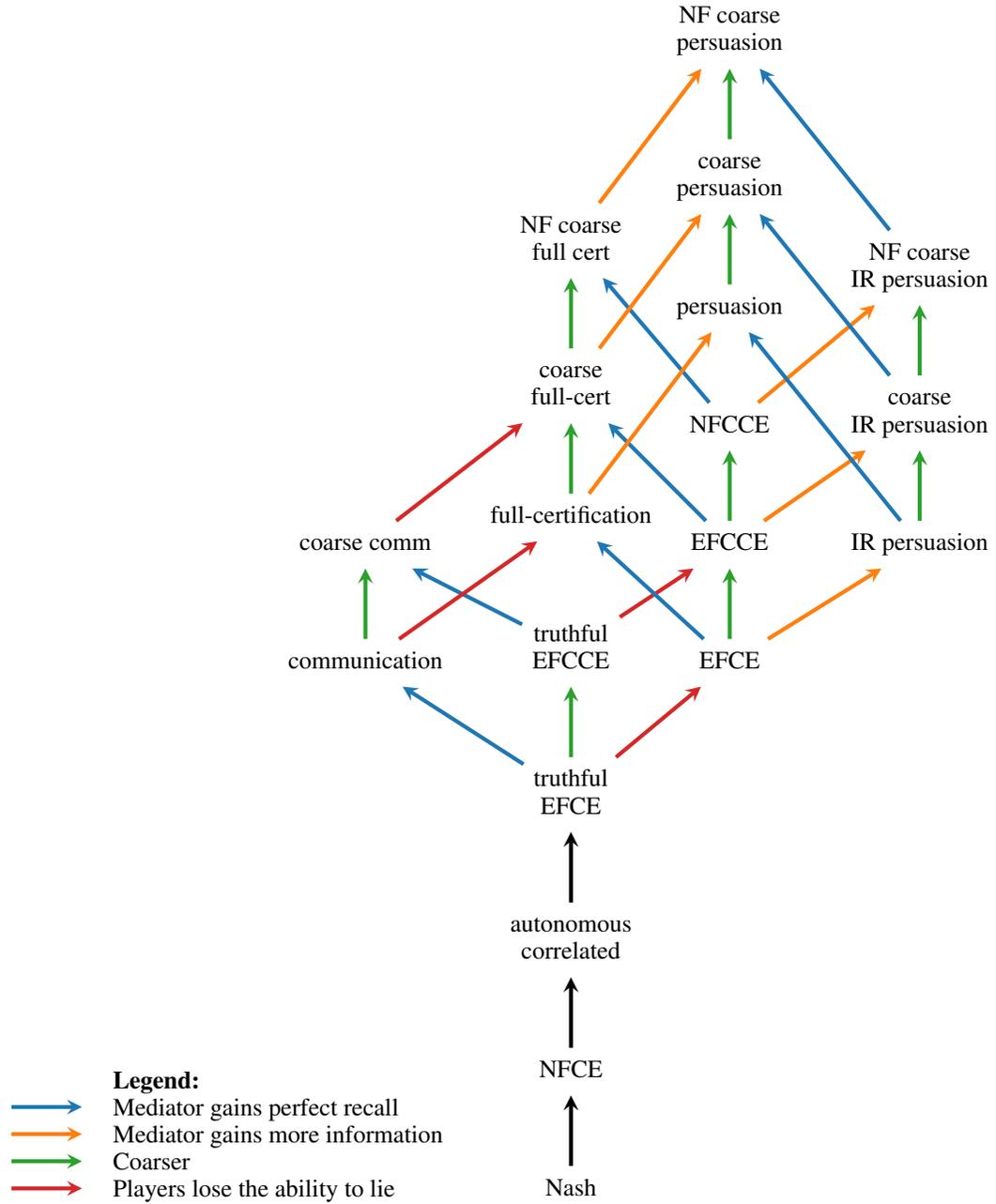
\begin{figure}[H]

\tikzset{
    every path/.style={-stealth,ultra thick},
    recall/.style={draw=p1color},
    recall2/.style={draw=p2color},
    coarseness/.style={draw=p3color},
    lying/.style={draw=p4color},
}

\newcommand{\arrow}[1]{
\begin{tikzpicture}
\node (a) [] {};
\node (b) [right=1cm of a] {};
\draw [#1] (a) to (b);
\end{tikzpicture}
}

\begin{center}
\begin{tikzpicture}[auto]
\node (nash) [] {\makecell{Nash}};
\node (nfce) [above= of nash] {\makecell{NFCE}};
\draw (nash) to (nfce);
\node (ace) [above= of nfce] {\makecell{autonomous \\ correlated}};
\draw (nfce) to (ace);

\node (tefce) [above= of ace] {\makecell{truthful \\ EFCE}};
\draw (ace) to (tefce);
\node (tefcce) [above= of tefce] {\makecell{truthful \\ EFCCE}};
\draw [coarseness] (tefce) to (tefcce);

\node (efce) [above right= of tefce] {\makecell{EFCE}};
\draw [lying] (tefce) to (efce);
\node (efcce) [above= of efce] {\makecell{EFCCE}};
\draw [coarseness] (efce) to (efcce);
\draw [lying] (tefcce) to (efcce);
\node (nfcce) [above= of efcce] {\makecell{NFCCE}};
\draw [coarseness] (efcce) to (nfcce);

\node (comm) [above left= of tefce] {\makecell{communication}};
\draw [recall] (tefce) to (comm);
\node (ccomm) [above= of comm] {\makecell{coarse comm}};
\draw [coarseness] (comm) to (ccomm);
\draw [recall] (tefcce) to (ccomm);

\node (cert) [above= of tefcce] {\makecell{full-certification}};
\draw [lying] (comm) to (cert);
\draw [recall] (efce) to (cert);
\node (ccert) [above= of cert] {\makecell{coarse \\ full-cert}};
\draw [coarseness] (cert) to (ccert);
\draw [lying] (ccomm) to (ccert);
\draw [recall] (efcce) to (ccert);
\node (nfcfce) [above= of ccert] {\makecell{NF coarse \\ full cert}};
\draw [coarseness] (ccert) to (nfcfce);
\draw [recall] (nfcce) to (nfcfce);

\node (sm) [above= of nfcce] {\makecell{persuasion}};
\draw [recall2] (cert) to (sm);
\node (csm) [above= of sm] {\makecell{coarse \\ persuasion}};
\draw [recall2] (ccert) to (csm);
\draw [coarseness] (sm) to (csm);
\node (nfcsm) [above= of csm] {\makecell{NF coarse \\ persuasion}};
\draw [coarseness] (csm) to (nfcsm);
\draw [recall2] (nfcfce) to (nfcsm);

\node (irp) [above right= of efce] {\makecell{IR persuasion}};
\draw [recall2] (efce) to (irp);
\node (cirp) [above= of irp] {\makecell{coarse \\ IR persuasion}};
\draw [recall2] (efcce) to (cirp);
\draw [coarseness] (irp) to (cirp);
\node (nfcirp) [above= of cirp] {\makecell{NF coarse \\ IR persuasion}};
\draw [coarseness] (cirp) to (nfcirp);
\draw [recall2] (nfcce) to (nfcirp);
\draw[recall] (irp) to (sm);
\draw[recall] (cirp) to (csm);
\draw[recall] (nfcirp) to (nfcsm);

\node (legend) [left=5cm of nash.south, anchor=south] {
\makecell[l]{
\arrow{draw=none} {\bf Legend:} \\
\arrow{recall} Mediator gains perfect recall \\
\arrow{recall2} Mediator gains more information \\
\arrow{coarseness} Coarser \\
\arrow{lying} Players lose the ability to lie
}
};
\end{tikzpicture}
\end{center}
\caption{Inclusion diagram for the equilibrium notions in \Cref{se:family} and a few others. 
{\em Autonomous correlated equilibria}~\cite{Forges86:Approach,Stengel08:Extensive} are equilibria in which the mediator cannot receive information from the players but still only gives recommendations one timestep at a time.}
\end{figure}
\end{document}